\documentclass[draft]{agujournal2018}
\usepackage{apacite}
\usepackage{multirow}
\usepackage{lscape}

\draftfalse
\journalname{JGR-Space Physics}

\newcommand{\solphys}{\textit{Solar Phys.}}
\newcommand{\nat}{\textit{Nature}}
\newcommand{\apj}{\textit{Astrophysical Journal}}
\newcommand{\jgr}{\textit{J. Geophys. Res.}}
\newcommand{\ssr}{\textit{Space Science Research}}
\newcommand{\aap}{\textit{Astronomy and Astrophysics}}
\newcommand{\apjl}{\textit{Astrophysical Journal Letters}}
\newcommand{\grl}{\textit{J. Geophys. Res. Lett.}}

\begin{document}
\title{Low geo-effectiveness of fast halo CMEs  related to the 12 X-class flares in 2002}

\authors{B. Schmieder\affil{1,2,3}\thanks{visitor in KASI}, R.-S. Kim\affil{2,4},  B. Grison\affil{5},  K. Bocchialini\affil{6},  R.-Y. Kwon\affil{2}, S. Poedts\affil{3,7}, P. D\'emoulin\affil{1}}

\affiliation{1}{LESIA, Observatoire de Paris, PSL Research University, CNRS Sorbonne Universit\'e, Univ. Paris 06, Univ. Paris Diderot, Sorbonne Paris Cit\'e, 5 place Jules Janssen, F-92195 Meudon, France}
\affiliation{2}{Korea Astronomy and Space Science Institute, 776 Daedeok-daero, Yuseong-gu, Daejeon 34055, Korea}
\affiliation{3}{Dept. of Mathematics, KU Leuven, Celestijnenlaan 200B, 3001 Leuven, Belgium}
\affiliation{4}{University of Science and Technology, 217 Gajeong-ro, Yuseong-gu, Daejeon, 34113, Korea}
\affiliation{5}{Department of Space Physics, Institute of Atmospheric Physics of the Czech Academy of Sciences, Bocni II, 1401, 141 00 Prague 4, Czech Republic}
\affiliation{6}{Institut d'Astrophysique Spatiale, Univ. Paris-Sud, CNRS, Universit\' e Paris-Saclay, B\^atiment 121, 91405 Orsay CEDEX, France}
\affiliation{7}{Group of Astrophysics, Institute of Physics, University of Maria Curie-Sk{\l}odowska, ul.\ Radziszewskiego 10, PL-20-031 Lublin, Poland}

\begin{keypoints}
  \item We study all the X-class flares and associated CMEs in 2002 (12 cases)
    \item None of the six     halo CMEs and shocks evoked strong geomagnetic disturbances 
      possibly due to the weak southward magnetic field components of the ICMEs.
    \item The favorable solar proxies (large  sunspot, 
   fast halo CME, ...) were  not sufficient to accurately forecast the geo-effectiveness of these CMEs.
  \end{keypoints}


\begin{abstract}
It is generally accepted that extreme space weather events tend to be related to strong flares and fast halo coronal mass ejections CMEs. In the present paper, we  carefully identify  the chain of events from the Sun to the Earth induced by all 12 X-class flares that occurred in 2002. In this  small sample, we find an unusual high rate (58\%) of solar sources with a  longitude larger than  74 degrees. Yet, all 12 X-class flares are associated with at least one CME. The fast halo CMEs (50\% ) are related to interplanetary CMEs (ICMEs)  at L1  and weak  Dst minimum values ($> -51\;$nT); while 5 (41\%) of the 12 X-class flares are related to solar proton events (SPE).\\
We conclude that: (i)   All twelve  analyzed solar events, even those associated with fast halo CMEs originating from the central disk region,   and those ICMEs and SPEs were not very geo-effective.
This unexpected result demonstrates that the  suggested events in the chain  (fast halo CME, X-class flares, central disk region, ICME, SPE) are not  infallible  proxies for geo-effectiveness. (ii) The low value of integrated and normalized southward component of the IMF ($B^*_z$)  may explain the low geo-effectiveness for this small sample. In fact,  $B^*_z$ is well correlated to the  weak Dst and low  auroral electrojet (AE) activity.  Hence, the only space weather impact at Earth in 2002 we can explain is based on $B^*_z$ at  L1.  

\end{abstract}


\section{Introduction}

It is generally {recognized} that extreme space weather events are often related to large, mature sunspot groups. Such turbulent and highly complex solar active regions (ARs) have a large amount of "free" (non-potential) magnetic energy, which  can be released during flares.  
Flares irradiate large amounts of thermal energy, up to $10^{33}\;$ergs \citep{Aulanier2013}. Intense X-class flares have irradiance values (E) exceeding $10^{-4}\;$W/$m^2$. They even affect the ionosphere due to their X-ray emission flux which can perturb the geo-localization systems \citep{Tsurutani2009}. 
Fast CMEs are usually associated with flares and are enormous eruptions of plasma (up to 10$^{13}- 10^{16}\;$g) and magnetic field into the interplanetary space at velocities up to several thousand km/s which interact with the solar wind and often also with the terrestrial magnetosphere \citep{Denig2015}. The kinetic energy of CMEs reaches up to 10$^{32}\;$ergs. Clearly, full front-side halo CMEs have more potential to arrive at the Earth a few days after their launch \citep{Kim2005}.
Magnetic reconnection during flare onset and shock waves driven by CMEs can both generate solar energetic particles (SEPs), {creating sporadic concentrations at Earth known as Solar Particle Events (SPEs) which are characterized by enhancements of several orders of magnitude above background levels, e.g.\ for protons energies vary from less than $1\;$MeV to over $1\;$GeV in extreme cases} \citep{Lario2004, Nitta2006, Cliver2007, Cane2010,Kim2015}.
CMEs surfing on the solar wind are called interplanetary CMEs (ICMEs). When they  are embedded in {  a strong magnetic field with a smooth rotation, and have a low plasma beta and proton temperature, they are called magnetic clouds (MC) \citep{Burlaga1981}.}
Consequently, {flare-related CMEs} are key in current space weather operational forecasting systems considering their potentially severe impact on Earth, {electric} power blackouts, and damage to spacecraft instrumentation. There are considerable research efforts underway to improve our understanding of {CMEs}  to provide feedback to operational space weather forecasting infrastructure \citep{Vourlidas2019}. The predicted time of occurrence of geo-effective events in the Earth environment depends on the effects studied:
 
\begin{itemize}
    \item The common transit time from Sun to Earth for CME/ICMEs and coronal hole high speed wind streams can be estimated up to 2 to 4 days \citep{Gopalswamy2007, Scolini2018}, i.e.\ only after this time the related effects such as geomagnetic storms, energetic {particle} enhancements, and geomagnetic induced currents (GIC) are measured at Earth \citep{Pirjola2013}.
    \item Impulsive SPEs (from a few keV) up to relativistic energies (few GeV), accelerated by magnetic reconnection driven processes during the impulsive phases of solar flares, have a transit time as low as  ten minutes while large gradual SPE events are produced by fast CME driven shock waves and can last a few hours or even days depending on their trajectories and their orientation in the interplanetary magnetic field (IMF) \citep{Zurbuchen2006}. SPEs can penetrate down to the upper atmosphere leading to solar array degradation, single event effects in electronics and potential health risks for manned spaceflight and air travelers  \citep{Lario2004}.
\end{itemize}

Many different methods exist to predict such events based on the determination of the initial speed, direction, shape and evolution of CMEs through the inner heliosphere and in the IMF  \citep{Cliver1990, Tsurutani1988, Tsurutani1992, Tsurutani2003}. The extreme space weather events which arrive { at  Earth} eight minutes after flares are not predictable without a good prediction of the occurrence of large flares. The largest 'extreme' events are exceptional events occurring once or even less than once during a solar cycle \citep{Wimmer2014, Denig2015}. Solar energetic particles of moderate strength \textcolor{blue}{occur} more frequent, as well as geomagnetic storms. Geomagnetic storms driven by ICMEs may be predicted, based on the CME characteristics, between two to four days before arriving in the Earth environment. It is difficult nevertheless to get a better forecasting accuracy than 50\% taking into account the different  CME parameters \citep{Vrsnak2005, Gopalswamy2007, Richardson2010, Bein2012, Jang2017}. Moreover, interaction of ICMEs and interplanetary structures, {\it e.g.}, high speed streams, are also reported to be drivers of severe geomagnetic activity \citep{Gonzalez1996, Wang2003, Xie2006, Lugaz2017}. Geomagnetic storms have been intensively studied by \citet{Cliver1990, Tsurutani1992}. 
The $B_z$ component of the IMF is well correlated to the storm Dst value \citep{Wu2002}. Moreover, ICMEs generating long periods of strong southward (negative) $B_z$  can interact with the magnetosphere magnetic  field  and initiate geomagnetic storms \citep{Tsurutani1988, Gosling1990}. This effect is even increasing with high speed solar wind or fast ICMEs.

{Statistical} approaches are important to detect associations and correlations in the chain of events from Sun to Earth. \citet{Richardson2010} listed 300 near Earth ICMEs and summarized their basic properties and their related geomagnetic effects.  \citet{Gopalswamy2007} listed 378 halo CMEs during solar cycle 23 and their geo-effectiveness (characterized by Dst$_{min}$). These authors confirmed  that halo CMEs originating  near the central meridian  have a better chance to be highly geo-effective. They also noted that the geoeffectiveness rate is particularly low in 1999 and 2002 (the beginning and the end of the maximum solar phase  of solar cycle 23). \citet{Cane2010} examined all the energetic particle events  occurring during solar cycle 23 and suggested that their main origins are flares with short or long duration depending on whether { they} were associated with CMEs or not. \citet{Cane2010, Dierck2015, Miteva2018, Matamoros2015} concentrated their studies on the SEP events during solar cycle 23. It was found that the probability of an SEP event occurring is higher for intense flares, high speed CMEs and sources with western longitude. It was also confirmed that there is a strong relationship between halo CMEs and SPEs \citep{Chandra2013}. \citet{Kim2015} classified the main SPEs occurring during solar cycle 23 in four categories according to  their onset times compared with the flare peak and  the delays existing between the  different channels of energy. \citet{Nieves2019} provided a full catalog of the ICMEs  observed between 1995 and 2015. These authors investigated the magnetic field configuration associated with 353 ICMEs in order to validate their flux-rope structure. The \citet{Jian2006a} catalog gathers most of the ICMEs identified during this time period.

However, most of the studies do not consider the whole chain of events from the solar surface to the Earth, {as this is rather difficult when starting from the solar source as opposed to starting from the ICME or its geo-effectiveness. Instead, they} focus merely on some parts of the chain or study only individual extreme events \citep{Tsurutani2003b, Denig2015,Kim2005,Kim2015, Schmieder2018}. Therefore, the physical chain of processes is usually not really studied as a whole. A scatter plot diagram between two events in the chain might give a higher probability of their correlation but this does not explain why they are related, i.e.\ what physical processes are behind their relation. Therefore, the physics of each of the phenomena in the chain is important and a deeper insight in this should {help  determining} the boundary conditions of numerical simulations. We think that we are still in the era of specific case studies leading to geo-effectiveness or not, trying to really understand why a chain of events starting from a flare-CME with all the favorable conditions does or does not end with a geo-effective event. Only a few rare studies are conducted to explain non-geo-effective events  \citep{Mays2015, Thalman2015}. As already mentioned, statistical approaches are important to detect associations and correlations in the Sun-Earth chain of events. However, they are often biased by the choices of (only) geo-effective events and the related solar events. Only a few works concern a not biased sample due to a careful selection of events. For example, \citet{Bocchialini2018} studied all the sudden storm commencements (SSCs) events occurring during one year without suppressing any event and following the chain of events from the Earth to the Sun. It was particularly difficult to relate the SSCs occurring at L1 to corresponding solar events and also to find the solar sources of the CMEs involved.

In the present paper we follow a new approach, not biased by the pre-selection of the flares, and we consider {\em all} the X-class flares occurring in 2002 and we trace the full chain of events from the Sun to the Earth.

The paper is organized as follows.
In Section 2 we first describe the catalogs from which we extracted the data that we used and then we address the methodology for following the chain of events from the Sun to the Earth.  The geo-effectiveness is discussed based on four different measurements: the Dst minimum, SPE flux values, the Auroral Electrojet (AE) maximum values, and SSCs.
 In Section 3, we  summarize the associations that we found between these phenomena  (Table~\ref{all_events}). Then we discuss the relationship between the characteristics of the flares, CMEs, ICMEs, and SPE events. In Section~4, we present in more detail the flare-CME chains of a few cases according to the longitude of their sources and particularly those for which all the parameters are favorable for being associated with geo-effectiveness. We try to find the {answer  to } why the geo-effectiveness rate was so unusually low in 2002 \citep{Gopalswamy2007}. This study should give insight to enable narrowing the space of parameters and ultimately lead to improved forecasts.

\section{Data and Methodology}
\subsection{Data}
The recent statistical study of \citet{Bocchialini2018} was dedicated to the chain of events from the Earth to the Sun by starting their research considering all the registered SSCs in 2002.
SSCs involve a sudden growth of the magnetic field strength at the Earth surface.
Furthermore, SSCs related to ICMEs are commonly interpreted as being due to the shock preceding the ICME sheath. For 2002, it was found that  the 36 SSCs that occurred were related to moderate (-100 nT $<$ Dst$_{min}$) $<$ -50 nT) or weak geomagnetic storms.{ On the other end of the chain, the SSCs were related to solar sources and only four of them correspond to a X-ray flare, i.e.\ about 30 \% (four among twelve).} We here revisit the data of 2002 with a different approach {in  which we  consider } all the X-class flares observed during one year (2002) {listed in Table~\ref{all_events}} and to follow the chain of events from the Sun to the Earth (until the magnetosphere). In the year 2002, at the end of the solar maximum period of solar cycle 23, 2727 flares (12 X-class, 219 M-class, and 2496 C-class flares) were registered by the GOES X-ray detector and listed by NOAA including their source region with their coordinates (\url{ftp://ftp.swpc.noaa.gov/pub/warehouse/2002}). 

In a first step, we are looking for the CMEs that could be related to those 12 X-class flares in the CDAW/LASCO CME online catalog  (\url{http://cdaw.gsfc.nasa.gov/CME\_list/index.html}) \citep{Yashiro2004}. 
{Those CMEs of which the  first appearance time was close to those of the 12 X-class flares from the CDAW list were selected. To check the association with the flares, we carefully compare the SOHO/LASCO images with SOHO/EIT running difference images (see an example in Figure \ref{case10}). {We determine EUV features such as coronal dimming, post eruption arcades or erupting filaments.} By determining whether the CME position and EUV features
have spatial and temporal closeness, we select the  related CMEs for each flare.} 

Then, we consider {the} solar energetic particles  (electrons, protons, ions){, mainly the solar proton events (SPE) listed in 2002 by the NOAA/Space Weather Prediction Center  (NOAA/SWPC)  (\url{https://umbra.nascom.nasa.gov/SPE/}). In this study SPEs are defined as an event where the number of incoming energetic ($>10\;$MeV) protons (Ip flux) exceeds 10 particles flux unit (pfu: proton number per {cm$^2$/sr/s}) at geosynchronous satellite altitudes. { We note that for some cases in the GOES curves in the low frequency range (($>10\;$MeV) there is a second maximum, higher than the first one which does not exist in the higher  frequency bands. The second maximum is interpreted as an energetic storm particle event (ESP). The particles are trapped in the ICME shock and  are locally accelerated in the vicinity of the Earth \citep{Makela2011}. Therefore we consider only the time of the first maximum as the peak time of SPE which  does not always correspond  to the peak time found in the catalogs  as we will show with  one  example in Section 3.} Five SPEs are related to the X-class flares, among the 19 SPEs  listed in the data base of solar energetic protons. 

We follow the events at L1 by using the OMNIWeb service   (\url{http://omniweb.gsfc.nasa.gov/}) and the AMDA service  (\url{http://amda.cdpp.eu}) to draw the records of all the  parameters (IMF magnitude, three IMF components, IMF inclination, Solar Wind plasma   density, bulk velocity, radial temperature, $\alpha$-particle to proton ratio). Statistical surveys of geomagnetic storms and their interplanetary causes become more routinely accessible  during solar cycles 23 and 24 due to the reliability of in-situ measurements in the interplanetary medium made by the two satellites WIND with the {\it Solar Wind Experiments} (SWE, \citet{Lepping1995}) and the {\it Advanced Composition Explorer} (ACE, \citet{Chiu1998,Stone1998}) with the magnetic field experiment (MAG) and plasma particle experiment (SWEPAM, \cite{McComas1998, Smith1998}).  These satellites  provide in situ measurements at L1  respectively on the interplanetary magnetic field strength and its components, the plasma velocity and density. 
{ Published catalogs \citep{Jian2006a, Richardson2010, Nieves2019} are considered for ICME identification.  {We adopt the method described in \citet{Bocchialini2018} to identify the ICMEs.} When the ICME shock velocity in catalogs is lower than the CME LASCO velocity and also lower than the ballistic velocity (the constant radial velocity that would permit to travel from the last CME LASCO observation  to L1 at the shock arrival time), we associate the ICME to the CME and we report ICME in Table \ref{all_events}.}
 {Long periods of intense southward IMF are known to be related to strong geo-effective events as day-side magnetic reconnection can occur. The peak value of $B_{z}<0$ or the mean($B_{z}<0$) can be misleading as these can be strongly negative even for very short $B_{z}<0$ periods. Moreover, the mean($B_{z}$) is close to 0 for events with equivalent periods of northward and  southward IMF even if it contains periods of long intense southward IMF. To estimate the IMF contribution to the geo-effectiveness we thus choose  $B_{z}^*$ as done in \citet{Bocchialini2018}.
$B_{z}^*$ is computed by the integration over time of the southward IMF $B_z$ component, divided by the total interval duration (including periods of northward IMF). }
 For ICMEs, $B_{z}^*$ is computed between the leading shock time at L1 and the time of min(Dst). 
 For the other events, $B_{z}^*$ is computed between the flare peak time plus one day and the time of min(Dst), and we report the value in brackets in Table \ref{all_events}.
$B_{z}^*$ values lower than $-5\;$nT are expected to be correlated to min(Dst).

Magnetic disturbances of the magnetosphere are studied from indices derived from measurements of station networks. Many different indices exist and correspond to different locations of the magnetometers (see \citet{Menvielle2011} for a detailed review).
Here we chose two indices, viz.\ the auroral electrojet (AE) for stations located in the auroral region, and the Dst index for stations in low latitudes, and a list of rapid magnetic variations (SSC, storm sudden commencements) published by the Ebre Observatory. SSC events are usually related to quick magnetosphere compressions. Moderate auroral activity is achieved for AE $>1000\;$nT. AE values above $1500\;$nT are related to an intense activity.
For the Dst we use the definition of \citet{Echer2008} and \citet{Gonzalez1994} to classify the geomagnetic storms as intense storm (Dst$_{min}$ $<-100\;$nT), moderate storm ($-100\;$nT $<$ Dst$_{min}$ $< -50\;$nT), and weak storm (-50 $<$ Dst$_{min}$). We report in Table \ref{all_events} the maximum of AE and the minimum of Dst observed for each event.
For ICME events we pick up Dst$_{min}$ and max(AE) values in the time interval starting from the leading shock time until the end of the magnetic ejecta observation time given in the previously mentioned catalogs. Other events (non-ICMEs) are a kind of reference sample. ICME geo-effectiveness can be estimated by comparing index values between ICME events and non-ICME events. We have to select index values of non-ICME events during an arbitrary period of time as there is by definition no ICME arrival time for these events. For the non-ICME events we consider a four day duration window starting from the flare peak time plus one day and we report the value in brackets. The time interval definition  explains the small differences between our values and the one presented in \citet{Bocchialini2018}.

\begin{landscape}
\begin{table}[!p]
\caption{Properties of the 12 X-class flares in 2002 and their related phenomena. Second column: date and time of the peak flux of the flare registered in GOES,  3rd column: X-ray class of the flare,  4th column: location of the source in heliographic coordinates, 5th column: active region (AR) in NOAA classification, 6th and 7th columns: coronal mass ejection  (CME), angular width (AW), speed (V$_{CME}$) { measured in the plane of sky   (radial speed for limb CME, expansion  speed for halo CME)},  8 -12  th columns: solar proton events (SPE): date/start time, peak time, $I_p$ (intensity), $\Delta$T (time between the  SPE start and  peak), $\delta t^d$ (time between the flare peak and the SPE start time), 13-14 th columns: ICME  (interplanetary coronal mass ejection), no (IP) no ICME but an interplanetary shock, $B_z^*$  (integrated normalized  B$_z$ component of IMF),  15-18 th columns:   SSC (Sudden Storm Commencement),   AE (auroral electrojet),  Dst$_{min}$ (when no ICME at L1 the values are in parentheses),  $\Delta$t (time between flare peak and  Dst minimum).}
\small
\tabcolsep=0.08cm
\centering
\begin{tabular}{cccccccccccccccccccc}
\hline 
\multirow{2}{*}{No.} & \multicolumn{4}{c}{Flare} & & \multicolumn{2}{c}{CME} & & \multicolumn{5}{c}{SPE$^b$} &  \multicolumn{2}{c}{ICME$^f$} &  \multicolumn{4}{c}{Earth's Disturbances}\\
\cline{2-5} \cline{7-8} \cline{10-14} \cline{17-20}
 & Peak Time& X-ray & Loc.$^a$ & AR & & AW & $V_{CME}$ & & Start Time & Peak Time & $I_{p}$ & $\Delta T^c$ & $\delta t^d$ & at & $B_z^*$ & SSC & AE & Dst$_{min}$ & $\Delta t^e$\\
 
 & (T1, UT) & class & & NOAA & & ($^\circ$) & (km/s) & & (T2, UT) & (T3, UT) & (pfu) & (hh:mm) & (hh:mm) & L1 & (nT)& & (nT) & (nT) & (day)\\
 \hline
 1 & 2 & 3 & 4 & 5 &&    6  &     7 &&      8 &    9&       10    &       11   &  12 & 13 & 14 & 15& 16& 17 & 18\\
 \hline
 1 & Apr-21 01:51 & X1.5 & S14W84 & 09906 & & 360 & 2393 & & Apr-21 02:25 & 11:25 & 2520 & 09:00 & 00:34 & ICME$^{J,N}$ & -2.6 & SSC11 & 1647 & -57 & 2.5\\
 2 & May-20 15:27 & X2.1 & S21E65 & 09961 & & 69 &  553 & &  { May-20              $^{C}$  } &--  &  --  &  --  &  $< $ 00:30 & no (IP) & (-1.4) &-- & (2018)& (-109) & --\\
 3 & Jul-03 02:13 & X1.5 & S20W51 & 10017 & &  73 &  265 & &       --     &       --     &  --  &  --  &  --  & no (IP) & (-1.4) &-- & (1054) & (-41) &3.2\\
 4 & Jul-15 20:08 & X3.0 & N19W01 & 10030 & & 360 & 1151 & & Jul-16 17:50 & 20:00 &  234 & 02:10 & 21:42 & ICME$^{J,R,N}$ & -0.9 & SSC20 & 697& -17 & 2.75\\
 5 & Jul-18 07:44 & X1.8 & N19W30 & 10030 & & 360 & 1099 & & Jul-19 10:50 & 15:15 &   13 & 04:25 & 27:06 & ICME$^{J,R}$ & -1.7 & SSC21 & 1180 & -36 & 2.90\\
 6 & Jul-20 21:30 & X3.3 & S13E95 & 10039 & & 360 & 1941 & & Jul-22 06:55 & 12:00 &   28  & 05:05 & 33:25 & no (IP) & (-1.5) & -- & (1176) & (-38) & 1.5\\
 7 & Jul-23 00:35 & X4.8 & S13E72 & 10039 & & 360 & 2285 & &        --    &       --     &  --  & -- & -- & ICME$^{J}$ & -1.4 &-- &  910 & -21 & 1.8\\
 8 & Aug-03 19:07 & X1.0 & S16W76 & 10039 & & 138 & 1150 & &    {    Aug-03         $^{C}$    } &  -- &  --  & -- & $<01:00$ & no & (0) & -- & (452) & (-31) & 1.5\\
 9 & Aug-21 05:34 & X1.0 & S12W51 & 10069 & &  66 &  268 & &        --    &       --     &  --  & -- & -- & no & (-0.2) & --& (408) & (-38) & 3.4\\
10 & Aug-24 01:12 & X3.1 & S02W81 & 10069 & & 360 & 1913 & & Aug-24 01:40 & 08:35 &  317 & 06:55 & 00:28 & ICME$^{J,N}$ & -3.4 & SSC26 & 1347 & -45 & 3.25\\
11 & Aug-30 13:29 & X1.5 & N15E74 & 10095 & &  57 &  254 & &        --    &       --     &  --  & -- & -- &  no & (-2.1) & -- & (1238) & (-109) & 2.75\\
12 & Oct-31 16:52 & X1.2 &   --   &   --  & &  26 &  160 & &        --    &       --     &  --  & -- & -- & no & (-2.9)& -- & (1254) &  (-75) & -- \\
\hline
\multicolumn{20}{l}{$^a$ -- : Backside event}\\
\multicolumn{20}{l}{$^b$ Higher energetic protons than 10 MeV {listed in SWPC- (C) additional SPE listed in \citet{Cane2010}}}\\
\multicolumn{20}{l}{$^c$ $\Delta T = T_{peak} - T_{start}$ = T3-T2}\\
\multicolumn{20}{l}{$^d$ $\delta t$ = T2-T1 }\\
\multicolumn{20}{l}{$^e$ $\Delta t = T_{Dst}$ - T1 }\\
\multicolumn{20}{l}{$^f$  ICME listed in the catalog of (J) \citet{Jian2006a}, (R) \citet{Richardson2010}, (N) \citet{Nieves2019}}\\
\end{tabular}
\label{all_events}
\end{table}
\end{landscape}

\subsection{Methodology for one example - Flare \#10 related effects}\label{methodo}
The methodology is as following. For example we consider the X3.1 flare occurring on August 24 in the NOAA AR10069 located S02W81 ({flare \#10 in Table \ref{all_events}}).{  The source region was identified { by the brightest region visible on the west limb in EIT images in 195$\;$\AA\ coinciding in time with the onset of a high speed (1913 km/s) halo CME listed in the CDAW catalog as shown in Figure \ref{case10}.} The flare erupted when the region was already close to the limb. }{ As we have no possibility to get magnetic information of sources at the limb, we look} at the region a few days earlier, namely on August 18. The region consists of a large sunspots group with a $\delta$ configuration (opposite following polarity {surrounding} the main leading polarity  - see Figure \ref{source} {bottom} panels). This magnetic complexity favors instability and magnetic reconnection leading to eruptions \citep{Aulanier2010}.

\begin{figure}[!b]
\centering
\includegraphics[width=30pc]{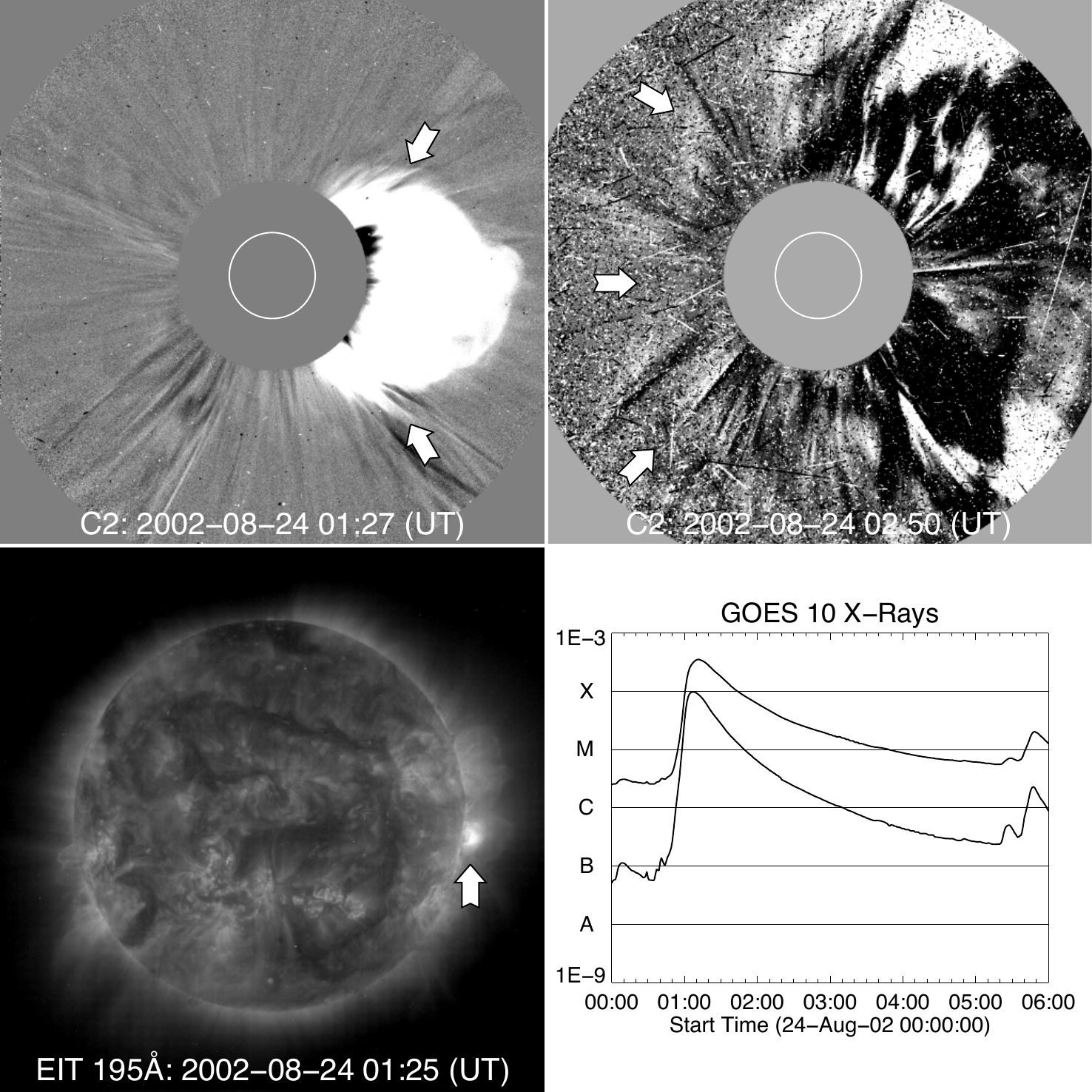}
\caption{Example of the halo CME source at  the West limb on August 24 2002 (flare \#10). 
Top panels: left, halo CME at 01:27 UT observed by the  LASCO/C2 coronograph, right, difference images of   the halo CME at 02:50 UT. The three white arrows indicate the front of the shock which appears even at the East limb completing the halo.
Bottom panels:  left, EIT image in 195 \AA\   showing the bright   flare   at the limb in AR 10069  with an  white arrow),  right, GOES X-rays record during the flare.}
\label{case10} 
\end{figure}

\begin{figure}[!p]
\centering
\includegraphics[width=30pc]{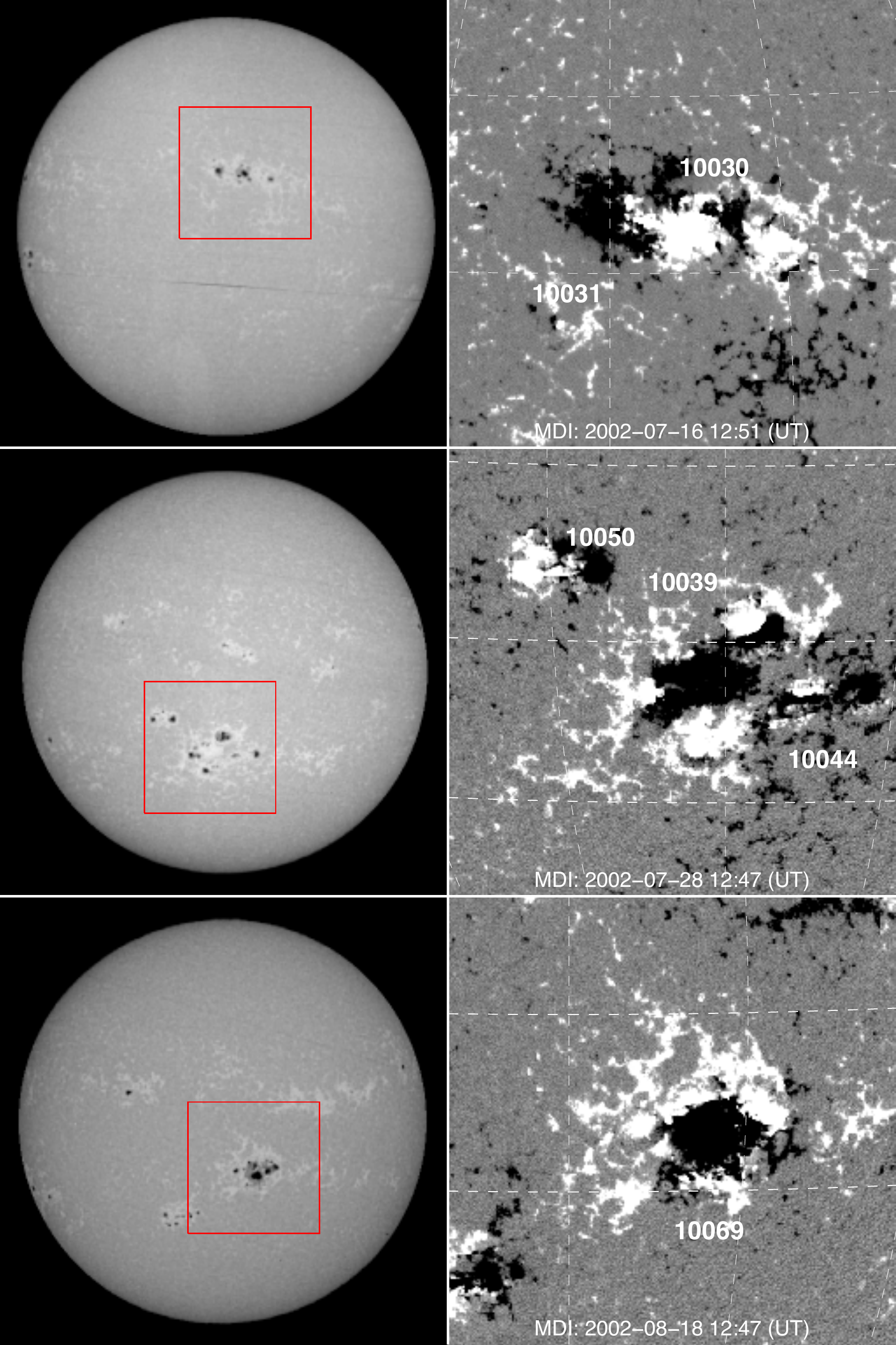}
\caption{Active regions (ARs) 
{ crossing the central meridian in July and August before or after  flare occurrence}, (left column: Meudon spectroheliograms in Ca K1v line),  and their complex magnetic field,  the field of view corresponds to the red box on the left,  (right column: MDI magnetograms from SOHO):  from top to bottom 
  - AR 10030 on 16 July between flares \#4 on July 15 and flare \#5 on July 18, -  AR  10039 on 28 July  between flares \#6 on July 20, \#7 on July 23  and \#8 on August 3, -  AR 10069 on 18 August  before flares \#9 on August  21, and  flare \#10 on August 24  (see Table \ref{all_events}).}
\label{source}
\end{figure}

{Then we look at the Wind  spacecraft data  in a time window   of four days after the CME launch.  Solar wind and IMF quantities are plotted between August 24 and 28 in Figure \ref{case10_l1}.
}
{Following the evolution of  the value of the $z-$component of the magnetic field at L1 (orthogonal component to the ecliptic plane) allows us 
to understand the possible reconnection in the IMF. $B_z< 0$ is a favorable condition for having reconnection between ICMEs and IMF \citep{Tsurutani1988}.}  

 \begin{figure}[!t]
	\centering
\includegraphics[width=40pc,angle=90]{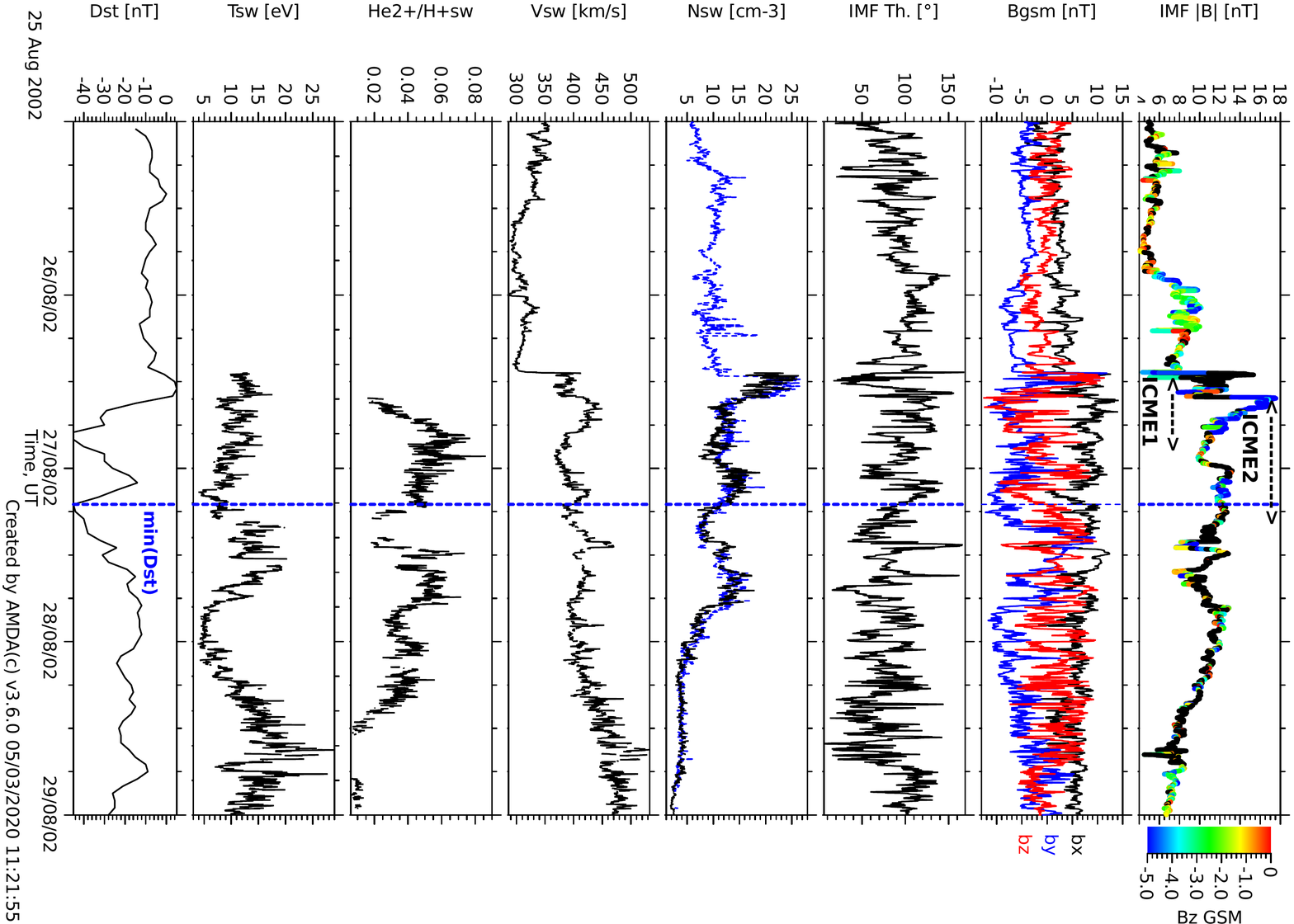}
	\caption{\small L1 observations (ACE data) starting from 24/08/2002 00:49 (detection of flare \#10) and during five days. IMF (Interplanetary Magnetic Field)  magnitude, three IMF components { and  IMF} deviation angle from $z_{GSM}$-axis
	are plotted in the three top panels (ACE/MAG data). The solar wind  plasma density, flow velocity, $\alpha$-particle to proton ratio, and  temperature are plotted in the next four panels (ACE/SWEPAM data). As there are some gaps in ACE data, density values from WIND/SWE instrument are plotted in blue.  In the bottom panel, Dst ground-based index is plotted and the {blue}  vertical dashed line indicates the Dst minimum related to flare/CME \#10.  { ICME1 and ICME2 are the ICMEs proposed by \citet{Jian2006a} and \citet{Nieves2019} respectively. }The time interval of { each  ICME} is indicated in the top panel. }
	\label{case10_l1}
\end{figure}

Two successive discontinuities are observed in the IMF magnitude (top panel of Figure \ref{case10_l1}) at 26 August at 10:40 UT and at 14:00 UT.
The first discontinuity is concomitant to the solar wind  kinetic pressure increase seen on both the solar wind density (fourth panel from the top) and velocity (fifth panel). The second IMF intensification is related to the end of the density pulse. It is worth noting that the IMF is strongly southward (second  and third panels) during several hours following the second discontinuity.
{ICMEs are reported during that period by \citet{Jian2006a} (ICME1) and \citet{Nieves2019} (ICME2). The magnetic obstacles identified in these studies are delimited by black arrows in top panel of Figure \ref{case10_l1}. A short magnetic ejecta is observed in ICME2 on the  27 August between 03:00 and 10:00 UT (low level of magnetic fluctuations and IMF  coherent rotation; see second panel).} 
This ejecta could be a {short duration ICME} with a positive helicity merging with the first one.  {In Figure \ref{case10_l1} we indicate by a vertical dotted line the Dst min corresponding to the ICMEs. We choose the time of the second Dst minimum  because  the ICMEs are still observed at L1 at that time and thus the Dst activity is probably ICME-related (see Table \ref{all_events}).}

{Furthermore, with the C2 coronagraph of LASCO,  
a partial CME is seen  around 20:50 UT on 23 August with a velocity of 860$\;$km/s, from very close parts in the AR before the launch of the  halo CME. This partial CME is caught by the halo CME. The limb location of the halo CME source can explain the poor aspect of the ICME observed at L1. The halo CME would correspond to the ICME between 27 August 03:00 to 10:00 UT {according to  the ballistic velocity analysis}. 

The ballistic velocity $v_{bal}$ ($\approx 650\;$km/s for this event) is the constant velocity that would permit an ICME to propagate from the solar corona at $t_{\odot{}}$ to L1 at the shock arrival time $t_{L1}$. The shock seen at L1 arrives with a velocity of $\approx 430\;$km/s ($v_{L1}$) and matches the  velocity criterion ($v_{bal}>v_{L1}$). As the CME velocity (1913$\;$km/s) is also larger than the ballistic  {velocity,} we can associate the CME related to flare \#10 to the observed ICME
We thus tag this event ICME in Table \ref{all_events}.}

The density  pulse obviously results in the onset of a magnetic storm. The Dst then reaches a minimum: -45$\;$nT between 19:00 and 20:00 UT on 26 August and again -45$\;$nT on 27 August at 03:00 UT corresponding to the two different events. $B_z$ is large and northward at the beginning (26 August at 11:00 UT). It turns quickly strongly southward at 15:00 UT but continues with strong rapid fluctuations until the  {end of the } second ejecta (ICME). It is consistent with the two successive minima of the Dst around -45$\;$nT.

We compute  $B_z^*$ (-3.4$\;$nT)  between the shock arrival time at L1 and the last Dst minimum (blue dot line, bottom panel of Figure \ref{case10_l1}). \citet{Bocchialini2018} noted that values {above} -5$\;$nT leads to moderate storms. {Therefore,  even if $B_z$ is strongly negative at the shock arrival time ($<-10\;$nT), if  it does not last long enough  (more than during half of the ICME duration)  it does not cause} a strong Dst decrease. The weak intensity of the storm is thus explained by the {sign of $B_z$, i.e.\ the } orientation of the IMF: the ICME magnetic field orientation at L1 is not optimum for causing large storm. We can also note a moderate auroral activity (AE$_{max}$ = 1347$\;$nT) when the ICME is passing through L1.

\begin{figure}[!b]
\centering
\includegraphics[width=30pc]{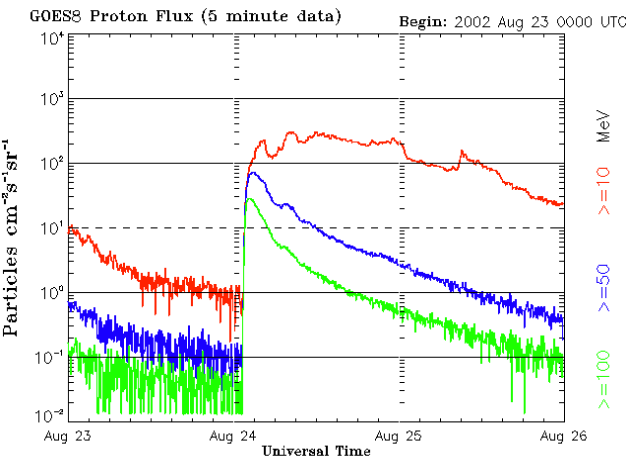}
\caption{Variation of SPE  on August 24  related to flare \#10.  In the low energy channel (red curve) several maxima occur during the gradual phase that are called Energetic Storm Particles (ESP).}
\label{spe_aug}
\end{figure}

Looking at  the SPE  data  listed in 2002 by NOAA/Space Weather Prediction Center and the GOES curves we  determine  approximately the SPE start  time (T2) and the SPE peak time (T3) avoiding the ESP {(Energetic Storm Particles)} events (Figure \ref{spe_aug}).  Their difference ($\delta$t) could  bring an  information about the location of the acceleration process. Flare/CME \#10 is related to an SPE with a short $\delta$t (28 minutes) after the flare peak time (T1), which means that the energetic particles could come directly from the flare reconnection site.{  However the CME is concomitant with the flare and has a very fast speed (more than 2000$\;$km/s). So it is difficult to disentangle if the SPEs come from the flare and/or the  CME/shock contributions \citep{Cane2010}.} SPEs are visible in the three frequency bands even in the more energetic one (Ip $> 100\;$MeV).
Such an event occurring in the central meridian would have been seriously geo-effective.

In summary, the flare-CME pair \#10 which has a solar source near the West limb (W81)  is related,  nevertheless, to an ICME at L1,  to a non negligible  Dst$_{min}$ (-45$\;$nT) and a SPE in the  Sun-Earth chain. The  tentative analysis  of this chain of  events   shows the  difficulty  to relate one event with {another}  one even if the times are consistent  with  the propagation velocities. 
We show  that  the conditions of the IMF  and the passage of previous CMEs are  important to interpret this event.

\section{Characteristics of the 12 X-class flares and the chain of related events}
\subsection{12 X-class flares and corresponding CMEs}

The twelve X-class flare-CME pairs and the chain of events from the Sun to the Earth  according to the methodology described above are listed in Table \ref{all_events}. The first five columns indicate the characteristics of the flares: number in the list, peak time, X-rays class, the coordinates of the solar source and  the NOAA AR number. The X-class flares are all originated in active regions with large sunspot areas and complex magnetic structures { mostly due to new emergence flux in the middle of the ARs} (Figure \ref{source}).  The flare occurrences are mainly clustered in one time period from 15 July to 30 August. During this time large sunspot areas consist of  $\beta, \gamma, \delta$ spots  (in the Hale nomenclature)  gathered in one or two ARs.

Indeed, large sunspot areas  with complex magnetic field are favored for X-class flares \citep{Aulanier2013,Schmieder2018}. It is in agreement with the good relationship found between the size of sunspots and the intensity of flares \citep{Sammis2000}. 
The source  of the 12 X-class flares  listed in  the fourth column of Table \ref{all_events} are located mainly  close to the limb (see Figure  \ref{location}).  Moreover, two flares  (flare  \#6 and flare \#12) have a backside solar source. 
However, the source of  flare \#6 could be identified by  bright loops prominent over the limb  before the AR 10039 passes on the disk (Figure \ref{source}). {Therefore, the flare source  is  defined as S13 E95 in the column 4 in Table \ref{all_events}.}

\begin{figure}[!t]
    \centering
     \includegraphics[width=30pc]{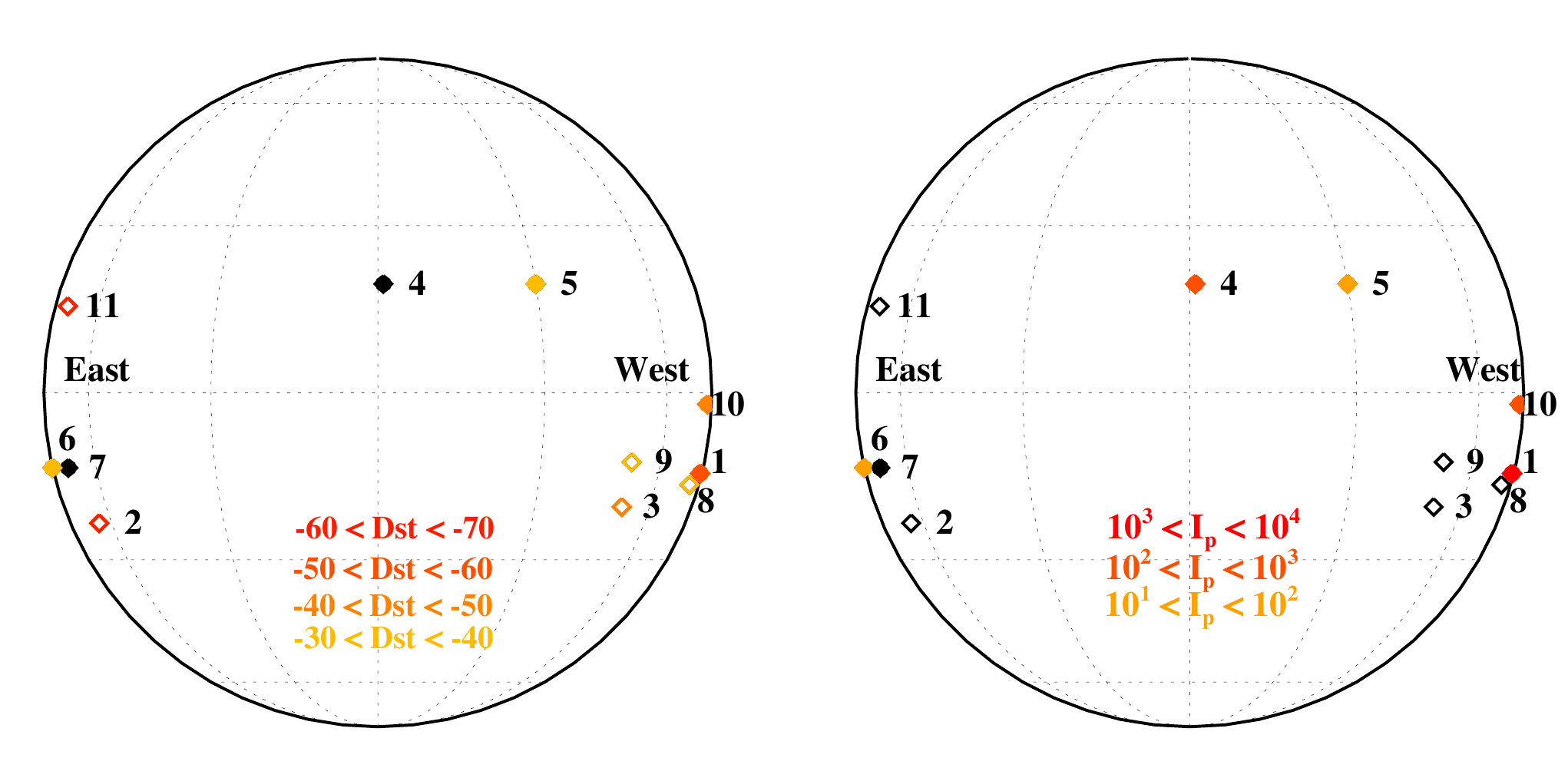}
    \caption{Left panel: longitudinal distribution of geomagnetic storm events associated with 12 X-class flares on 2002. The red- orange- yellow diamonds indicate the flare  sources associated with  -30 nT  $<$ Dst $<$ -60 nT.  The full  diamonds are associated with full halo CMEs in both panels. Right panel: Longitudinal distribution of the  SPEs  associated  with X-class flares on 2002 (red- orange- yellow diamonds for $10^1$  pfu $< $ I$_p$  $<$ 10$^3$ pfu).  The black diamonds are the location of the flare sources not associated to any SPE.}
    \label{location}
\end{figure}

\begin{figure}[!b]
\centering
\includegraphics[width=30pc]{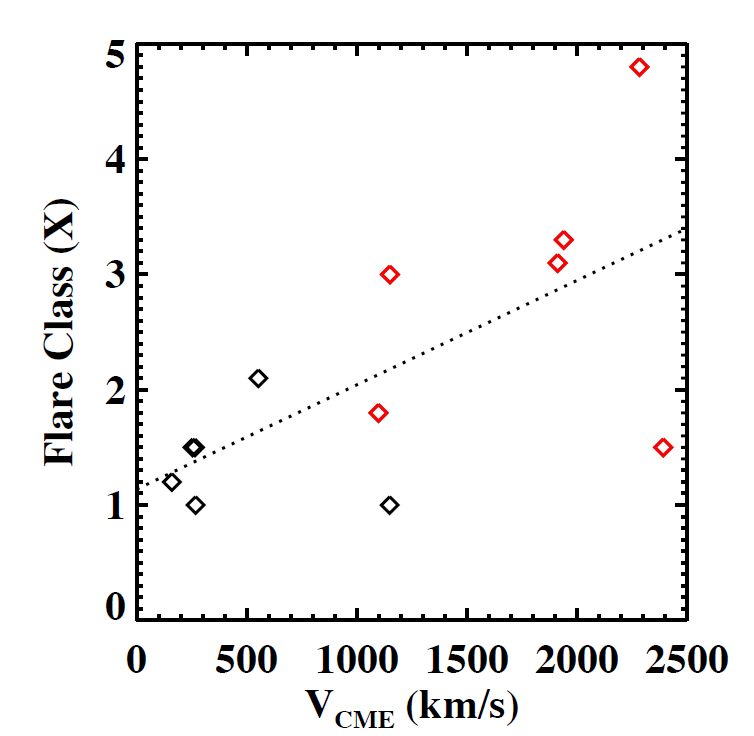}
\caption{Relationship of the 12  X class flares and CMEs  in 2002. 
 V$_{CME}$ is the speed of the CME (see Table 1). The red diamonds  correspond to   halo CMEs, the black  diamonds correspond  to non halo CMEs.}
\label{flare_cme}
\end{figure}

The 6th and 7th column concern the CME characteristics:  width and velocity. We note that all the X-class flares are associated with a CME: half of them  with a halo CME with a velocity between 1000$\;$km/s and  2000$\;$km/s. In this context we will consider  the { X-class  flare and its associated CME  as a global  event, in our study named a flare-CME pair.}  The flare and CME relationship  is shown in  Figure \ref{flare_cme}. The flare  flux is relatively well correlated to  the CME speed and with the CME width as well.
This confirms previous, more extensive statistical studies  associating flare flux with CMEs \citep{Hundhausen1997, Moon2002, Yashiro2006,Gopalswamy2009}.  We are in agreement with the identification  of \citet{Yashiro2004}, except for one flare (flare \#12 in our flare list), which was considered as a  CME-less flare   \citep{Klein2010}.  Finally, all the 12 X-class flares are associated with at least one CME (see Table \ref{all_events}).

 \subsection{Mass and kinetic energy of CME}
 
 {The intensities in coronagraph images result from Thomson-scattered light by coronal electrons (e.g., \citet{Minnaert1930}), and thus the number of electrons can be obtained from the observations by considering the Thomson scattering theory (e.g., \citet{Billings1966}). While the mass estimation also depends on the 3D geometry of the source (e.g., \citet{Vourlidas2006}), the CME mass has been determined under the assumption that all mass lies on the plane-of-the-sky \citep{Vourlidas2000}. After the multiple viewing perspective observations became available by STEREO in late 2006, there have been attempts to correct the CME mass with the 3D geometry \citep{Colaninno2009,Carley2012,Feng2013}. Due to the absence of the STEREO observations in our study, the CME mass has been derived under the plane-of-the-sky assumption.}

{We use the LASCO C2 images subtracted from a pre-event image to remove the background emission including the noise, and the residual was used for the calculation of the CME mass. The mass was derived from all images taken at times when the CME body passes through the height range from 2 to 6 solar radii, and the maximum values are listed in Table 2. To calculate the mass from C2 images, we used an IDL procedure, 'calc\_cme\_mass.pro' that can be found from the Solar Software. Note that the mass varies with time because of the propagation and the expansion as they are passing through the image plane, and the CME bodies became larger than the C2 FOV. In this regard, the calculated mass is not the total mass of the whole CME body, but the maximum value could be a proxy of it owing to conservation of mass. Once the mass is determined, the kinetic energy can be computed with the CME speed listed in Table 1. We note that all the halo CMEs have a kinetic energy larger than 10$^{32}\;$ergs except the CMEs \#4 and \#5, which have a slight lower energy around 10$^{31}\;$ergs. The non halo CMEs have a kinetic energy, one to three orders lower than the halo CMEs.

}

{The CME kinetic energy is a lower limit as the real CME velocity could be  larger than the apparent one,  due to projection effects.}

\begin{table}
    \caption{Characteristics of the CMEs, halo and partial related to the X-flares}
    \begin{tabular}{c|l|l|l|l}
       Number    & time of CME  & CME shape &  mass of CME & kinetic energy  \\
         flare & UT && [g]& [ergs]\\
          \hline
         1& 21 Apr. 01:27 & halo & 1 $\times$ 10$^{16}$ &3 $\times$ 10$^{32}$\\
         2& 20 May 15:25 & & 3 $\times$ 10$^{15}$& 5 $\times$ 10$^{30}$\\
         3& 03 July 02:10 & & 7 $\times$ 10$^{14}$& 3 $\times$ 10$^{29}$\\
         4    & 15 July 21:30 &halo& 6 $\times$ 10$^{15}$ &4 $\times$ 10$^{31}$\\
         5 &18 July 07:49 & halo& 1 $\times$ 10$^{15}$ &9 $\times$ 10$^{30}$  \\
         6 & 20 July  21:30 &halo& 9 $\times$ 10$^{15}$ & 2 $\times$ 10$^{32}$ \\
         7& 22 July 18:30  &halo& 5 $\times$ 10$^{15}$ &1 $\times$ 10$^{32}$\\
         8& 03 Aug. 18:59 & & 2 $\times$ 10$^{15}$ & 1 $\times$ 10$^{31}$\\
         9&  21 Aug. 05:28& & 2 $\times$ 10$^{15}$ & 9 $\times$ 10$^{29}$\\
         10 & 24 Aug. 00:49&halo & 2 $\times$ 10$^{16}$ & 3 $\times$ 10$^{32}$\\
         11& 30 Aug. 12:47& & 5 $\times$ 10$^{14}$ & 1 $\times$ 10$^{29}$\\
         12 & 31 Oct. 16:47&& 7 $\times$ 10$^{14}$ & 9 $\times$ 10$^{28}$\\
\end{tabular}
\label{tab:energy}
\end{table}

\subsection {Solar energetic particles}
Columns 8 through 12  of Table \ref{all_events} concern the SPE characteristics: start date and time, peak time, intensity, $\Delta$T (time between the start of the SPE and the peak), and  $\delta$t  (duration time between flare peak and start of the SPE).
$\Delta$T gives an estimate of the duration of the rising phase of the  phenomena (between 2 to 9 h).  $\delta$t  denotes the traveling time of the ICME before acceleration of particles. These  are  all approximate values and concern the low frequency band of energetic particles ($> 10\;$MeV).
Only five  flare - CME pairs  are related to SPEs. In Table \ref{all_events} we list the flux of these five SPE events in pfu.  We identify a new SPE event not listed in the  NOAA/SWPC data base: it concerns flare \#6 (see Table \ref{all_events}).
The pfu values are weak for most of them but there are not negligible with a highest value of 2520 pfu for the flare/CME \#1. 
Indeed, commonly only one SPE event  with a flux  $>$ 800 pfu occurs per year \citep{Klein2011}.
The maximum value that has been measured ever was 40000 pfu (24/03/1991) during the solar cycle 22 and  6300 pfu (07/03/2012) during the solar cycle 24. 

{ \citet{Cane2010} found 40 SPEs  in 2002  with {six of them related to an X-class flare. Four of them are listed  in  the SWPC catalogue and  two  of them  (flares \#2 and \#8 - Table \ref{all_events}) are not. The definition of SPE events in Cane's paper is based on the intensity of  protons  in the domain 25 to 30$\;$Mev. However,  it is difficult to identify an enhancement of the  flux  in the low energy band around 10$\;$Mev (GOES curves). The Ip is too  small to be listed in  the SWPC catalogue.   Therefore, we do   not use these two events in  the following discussion.}  However, we find one more   SPE event related to flare \#5. The occurrence of   a large type III, seen in WIND confirms our identification as an impulsive event.}
Long duration events or gradual events indicate that energetic particles are  accelerated  during a long period of time in the heliosphere, corresponding to the travelling time of the  ICME toward  the magnetosphere \citep{Tsurutani2009}. There is  a slight correlation between CME speed and  $\delta$t and with the SPE flux (Figure \ref{flare_spe}).  Short  $\delta$t  corresponds to fast CMEs.
 SPEs  are often associated with the local acceleration of the particles by interplanetary shocks.

\begin{figure}[!t]
\centering
\includegraphics[width=30pc]{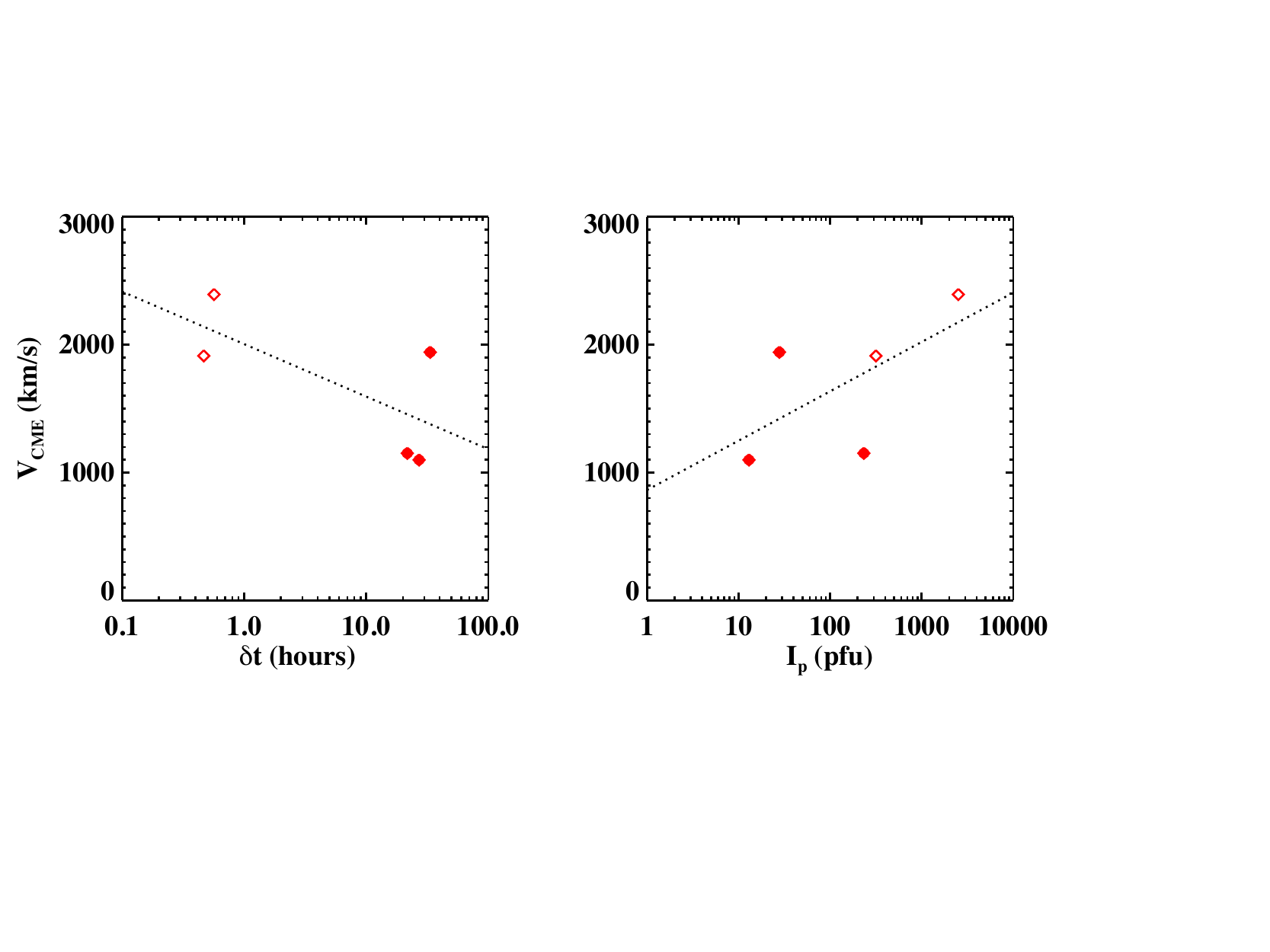}
\caption{Relationship of  X-class  flare-elated CME speed and solar proton events (SPE) in 2002. The energy Ip is in pfu, $\delta$T is  the time difference between the flare peak and the SPE start times. 
The full/empty red diamonds correspond to  long/short  SPE-flare delays. } \label{flare_spe}
\end{figure}

\subsection{ICMEs' arrival and geomagnetic indices}
Columns 13-14  of Table \ref{all_events} gather the L1 observations related to each {flare-CME pair}. Three categories of events are listed in column13: 
{"ICME" stands for clear ICME signature related to halo CME,  "no" stands for absence of ICME signature, and ''no (IP)" stands for no ICME but with the detection of an interplanetary shock (IP)  signature probably not related to the halo CME.
We could identify unambiguously five ICMEs (\#1, \#4, \#5, \#7, and \#10), all in agreement with previous catalogs 
\citep{Jian2006a,Nieves2019,Richardson2010},
three ''no (IP) cases (\#2, \#3, and \#6)   with a velocity at L1 lower than the ballistic velocity and 
four  "no"  cases  (\#8, \#9, \#11 and \#12),} with no time-consistent association when performing  the velocity comparison (CME velocity, solar wind velocity, and ballistic velocity). Among these four cases of  flare-CME pairs the associated  CMEs are slow and they are not observed at distance larger than 10 R$_\odot$.

The time is not in favour with the association with the CME because 
the ballistic velocity is lower than the velocity at the Sun and at L1. 
{Finally in column 14, we indicate  the southward IMF component (B$_z<0$) integrated and normalized value over time ($B_{z}^*$). In \citet{Bocchialini2018} $B_{z}^*$ above -5$\;$nT were associated with small Dst depression (low geo-effectivity). We can note that the minimum $B_{z}^*$ value is -3.4$\;$nT (\#10) indicative that no ICME is   expected to be strongly geo-effective.}

Columns 15-18 in Table \ref{all_events} indicate the Earth's disturbances: 
SSC, AE, Dst, and the delay  $\Delta t$ between the peak flare time and the minimum of the Dst. We recap that Dst values in parenthesis in Table \ref{all_events} are not attributed to the considered {flare-CME pair.}
This illustrates the utility of performing a detailed association between the Earth and the Sun and not taking a fixed time window.
The overall geo-effectiveness based on min(Dst) and max(AE) values consecutive to ICMEs is rather low. 
{The four SSCs   are associated with  four X-rays flare-CME pairs  having also  a SPE signature  (Table \ref{all_events}). They
display an enhanced auroral activity shown by  an important AE index.  A fifth  flare-CME pair 
(\#6) is also associated with SPEs. However, this pair  is not associated  with any SSC neither with a  Dst minimum.  
This could be explained  by the backside  source location of the flare. }
In column 18 of Table \ref{all_events} we see that the  disturbances arrival time   $\Delta t$  is between 1.5 and 3.4 days in the environment of the Earth. This leads to   a mean value of  the time window of 2.6 days.

  {In conclusion, the analysis of the Sun-Earth  chain of all these events shows that there is not a  significant  relationship between 
 Dst values and flare-CME characteristics (speed and width)  (Table \ref{all_events}
 and Figure \ref{location}).
  It is worth noting that none of the fast halo CMEs in our sample causes a strong Dst depression, despite they display the most relevant proxies for geo-effective event predictions. The number of our  events is certainly too small to make a  strong statistical conclusion.
  The source longitude is certainly a cause of this anomaly.}

\begin{figure}[ht]
\centering
\includegraphics[width=30pc]{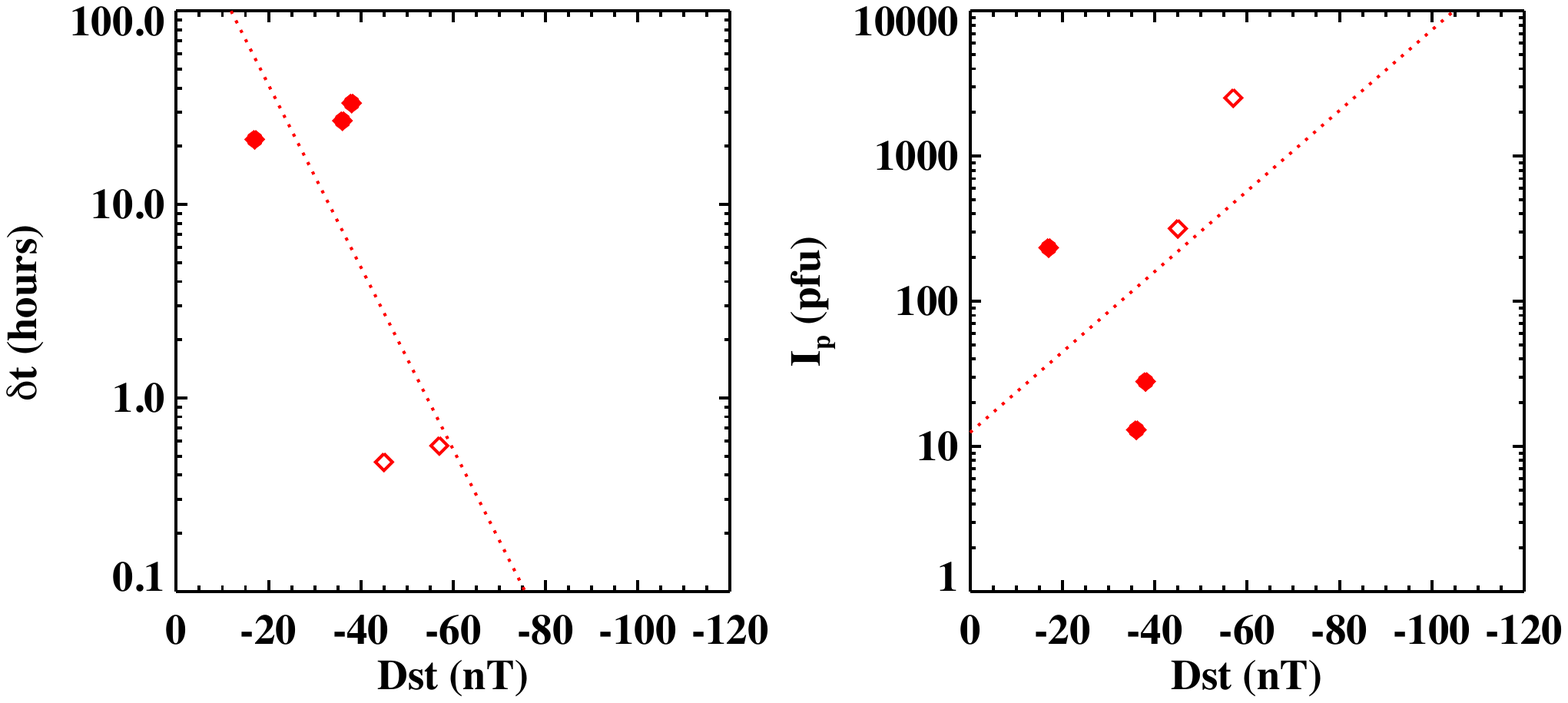}
\caption{Relationship between SPE characteristics  ($\delta$(t), I$_p$ flux) and Dst.
The full/empty red diamonds correspond to  long/short  SPE-flare delays.
}
\label{sep_dst}
\end{figure}

\begin{figure}[ht]
\centering
\includegraphics[width=30pc]{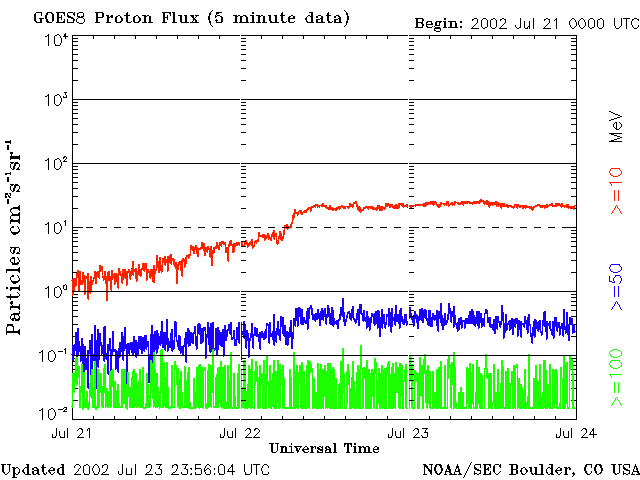}
\caption{Variation of SPE  between July 22 and 27  related to flare \#6. }
\label{spe_jul}
\end{figure}

\begin{figure}[ht]
\centering
\includegraphics[width=30pc]{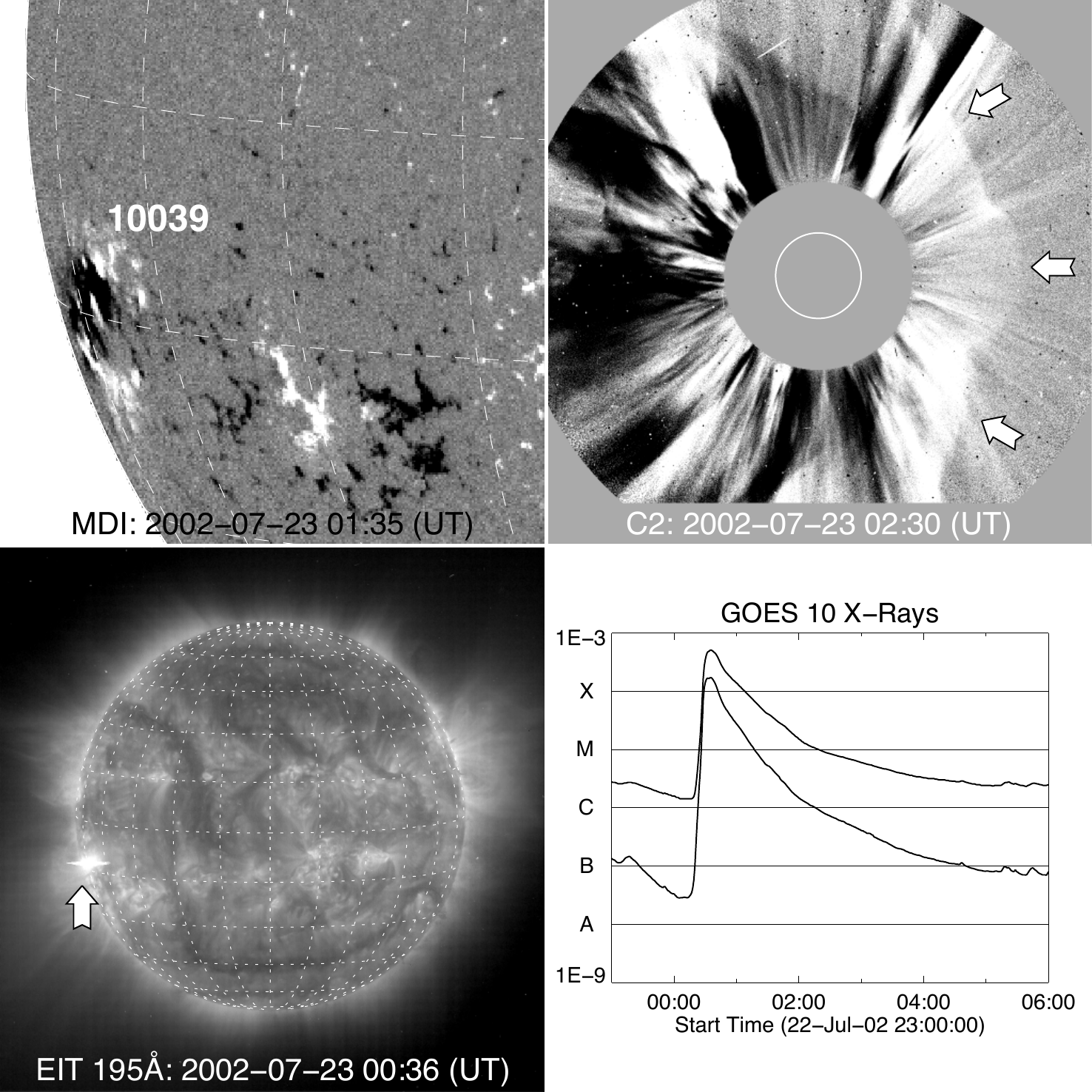}
\caption{Example of the  halo CME source in the East  on July 23 2002 and its solar source (flare \#7).  Top panels, left: MDI magnetogram of the flaring active region (AR 10039), 
right: difference image of the LASCO/C2  coronagraph showing the halo CME and the shock front on the West side,  indicated by white arrows.
Bottom panels, left: EIT image in 195 \AA\ showing the bright flare (see arrow),
right: GOES X-ray record.}
\label{case7}
\end{figure}

\begin{figure}[ht]
	\centering
	\includegraphics[width=30pc]{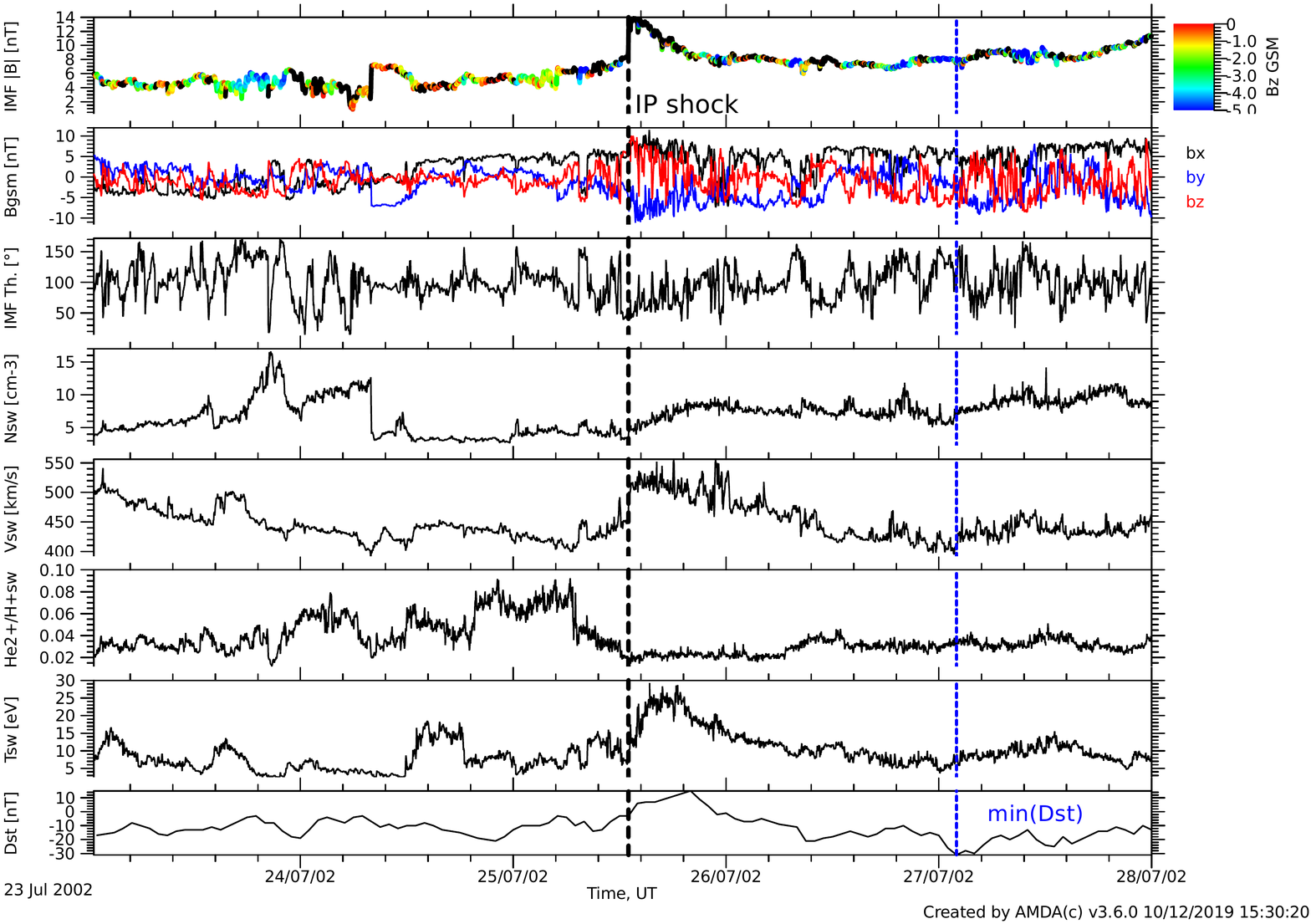}
	\caption{\small L1 and ground-based observations starting from the 23/07/2002 (time detection of flare \#7) and during 5 days. IMF magnitude, the three IMF components and IMF inclination are plotted in the three top panels (ACE/MAG data). The SW plasma density, flow velocity, radial temperature, and $\alpha$-particle to proton ratio are plotted in the next four panels (ACE/SWEPAM data). In the bottom panel, Dst ground-based index is plotted.  { the vertical dashed black  line indicates the interplanetary shock,  the vertical dotted blue line indicates the Dst minimum.}
	}
	\label{case7_L1}
\end{figure}

\subsection{Flare-CME source longitude and geo-effectiveness}
The longitudinal location  of solar flare-CME source is commonly  a good indicator to predict its  geo-effectiveness. 
 It is well known that limb events are less geo-effective than those originating around the central meridian \citep{Webb2002,Richardson2010,Cid2012,Lee2014}. However, it is  not 100\%  reliable \citep{Webb2002,Vasanth2015}. 
 {In \citet{Cid2012}, limb halo CMEs refer to a source longitude $>80\;$degrees.  We here adopt the same definition. }
We have already mentioned  in section 3.1 the  large proportion of our events 
having a source close to the limb (limb events) {(Figure  \ref{location})}.  It is an unusual high rate of limb sources in 2002 and it may explain the low geo-effectiveness of our sample. 

We compare the  longitudinal distribution of the source regions of { our sample limited to the  12  X-class flare-CME events  in 2002 } to  the longitudinal distribution  of the source regions of the CMEs related to all the   geomagnetic storms occurring in 2002 found in \citet{Bocchialini2018}: the percentage of sources close to the limb is much higher  for our sample (7/12 - 58 \%)  than the  mean value  (10/26 - 28 \%) found  in the sample of  \citet{Bocchialini2018}.
Dst-based geo-effectiveness of all the 12 X-class flares are weak, with only one moderate storm  (flare \#1) (Table \ref{all_events}, Figure  \ref{location}).
According to \citet{Gopalswamy2007} and \citet{Cid2012} such limb events (7 among 12 events) cannot 
lead to intense geomagnetic events, only moderate or weak events are expected. It is consistent with what we have observed. 

SPE   events  depend  also on the  West-East longitude of the solar sources. Among  
 the five SPE   events   that are listed in Table \ref{all_events} four of them have a West source.
 Particles coming from a West  source have  stronger flux  than those coming from the East due to the systematic eastward deflection of ICMEs when they interact with  the Parker spiral interplanetary magnetic field {\citep{Parker1958,Zurbuchen2006}}. 
  This is why the longitude of our  SPE  event sources  corresponds   mainly to the  West side of the Sun (Figure  \ref{location}).

{Figure  \ref{location} summarizes the main results of the  flare source longitude. For  the ICME subset there is  a relationship between  the flare  source longitude and Dst minimum values.} 
{However, the sample is too small to draw  a robust conclusion.} We note also  a   trend for AE  to occur  for  West side  sources.
More sources of X-ray flare-CME pairs are located in the West part (70 \% from the West part similar to the 61\% found  in \citet{Bocchialini2018}). 
This is  also in agreement with  the     statistics studies of large number of events \citep{Richardson2010}. 
From our (small) sample of   12 flare-CME pairs  we have the following classification:
\begin{itemize}
    \item four sources are on the East side (flares \#2, \#6, \#7, \#11), two are related to  an ICME.
    \item  five  sources are on the West side (flares \#1,\#3, \#8, \#9,   \#10), two are related to an ICME.
    \item two sources are front side (flares \#4, \#5) and are related to  ICME. 
    \item one source location is backside (flare \# 12).
\end{itemize}

Seven flare-CME pairs   have  no ICME signature at L1. None of them are in the disk central region:
four of them (\#2, \#6, \#7, \#11) have  a solar source  in the far East disk side, two of them  (\#8, \#9) in the far  West side, and one (\#12) has a source on the backside.
Moreover, six of these flare-CME pairs (all but \#6) are associated with narrow  CMEs.
The angular width  of these CMEs (AW in the 6th column in Table \ref{all_events}) varies  between   57  and  140 degrees.
The  longitudinal location of their solar sources  and the narrow width of the CMEs could be  good reasons why they have no signatures at L1. The  CMEs pass far away  from the Sun-Earth direction. 

In conclusion, we note the the source longitude is an important parameter for the geo-effectiveness of the events (e.g.\ Dst, SPE). 
{The  flare-CME   subset related to  ICME or  SPE 
intersects  with the halo CMEs subset of our events.
  The  limb source longitude of the CMEs
   can explain the poor geo-effectiveness measured in 2002, the  ICMEs avoiding the Sun Earth axis.  However, it does not explain the cases of the  front-side CME source events for which  the weak $B_z$ component of the IMF  cannot be predicted 
   (column 14 in Table \ref{all_events}).  The full chain of the events related to  the    six X-class flare associated with   halo CMEs  is discussed in details  according to their solar source  longitude and the IMF configuration  in the Annex.} \\
 
\section{Discussion of the results}
\subsection{Summary of the study of the 12 X-class flares}
In the present paper we focus on the full set of the 12 X-class flares observed in 2002, their solar sources, the  accompanying CMEs, and their related geo-effective events (indicated by the Dst index, the AE index, and the SPEs fluxes, Table~\ref{all_events}). This study tries to answer a question that became pertinent after the paper of \citet{Bocchialini2018} in which only four X-flares among the twelve 2002 X-class flares have been related to geo-effectiveness defined through the Sudden Storm Commencements (SSCs).  All the 12 X-class flares analyzed here are associated with at least one CME, and six of them even with fast halo CMEs
($1000\;$km/s $< $V $< 2000\;$km/s), while the other six CMEs are characterized by a slow propagation speed and a narrow width CME. We follow the fast CMEs as they propagated through the heliosphere and search for their signatures at L1 in a reasonable time window (1 to 4 days after onset). We find that the average arrival time of the ICMEs and IPs is 2.6 days for our small sample. Five flare-CME pairs among the six with fast halo CMEs could be related to ICMEs
producing  geomagnetic storms, detected by a minimum in the Dst index.

{ We identify the solar sources of the X-rays  flare-CME pairs and find an exceptionally high rate of sources occurring near the limb (more than 58\%).  Statistically it is  known that CME limb source  is not a favorable location for producing intense geomagnetic storms \citep{Gopalswamy2009,Cid2012}. In our sample, all the Dst minima are very weak, and only one flare-CME (\#1) is related to a moderate geomagnetic storm (min Dst $< - 50\;$nT).
The flare duration seems to be an important parameter in the chain of flare to geo-effectiveness since we  
 find a general trend for Dst, AE and Ip to be larger with longer duration flares. }

Moreover,  six cases among our sample display all most relevant proxies for geo-effective event predictions, e.g.\  fast halo CMEs.

For the five of them followed by ICMEs, the shock arrival times observed at L1 are consistent with their ballistic (constant) velocity. 
\citet{Bocchialini2018} showed that this is a relevant parameter to test the ICME/CME association.
 Among this set of these six flare-CME pairs  we identify four cases with a limb source. The associated  ICMEs are  identified among  complex fluctuations of the magnetic field due to shocks and sheaths in front of the ICMEs leading to a small value of B$_z^*$. {$B_{z}^*$ is computed by the integration over time of the IMF $B_z$ component, divided by the total interval duration. Then, even limb events could be detected at L1.}
The low B$_z^*$ explains their low impact on the magnetosphere.

Considering the SPEs, in our sample of twelve X-rays  flares,  five flares  (flares \#1, \#4, \#5, \#6, and \#10) are associated   with an SPE event.  They are also 
  associated with an ICME.
  The percentage of X-class flares related to SPE events associated with an ICME is here thus 40\% (5/12). In spite of our small sample, this is a similar percentage than the one found by \citet{Miteva2018} for the whole solar cycle 23. The corresponding sources of the SPEs are located at the West side or at disk center,  regions which are effectively favorable for the propagation of accelerated particles. Only one X-class flare (namely flare \#1) among the twelve is accompanied by a large SPE flux  (2520$\;$pfu). 
  For the SPEs, we confirm the conclusion of \citet{Klein2011}: these authors found that during 10 years (1996-2006) among 69 GOES X-class flares only 10 SPE events (one per year)  have a proton number larger than 800$\;$pfu. Furthermore, we note that we could identify two more SPE events for 2002 not discussed in \citet{Klein2011}, namely related to flares  \#5 and \#6 which occurred in July.  More generally, all the five SPEs  found are  related to X-class  flares with halo CMEs and to SSCs (except flare \#6 with no relation to an SSC). 

All the SPE delays between flare peaks and SPE start times,  are in good agreement with \citet{Cane2010}.  The former authors interpreted the long delays  by "a slow rise of the event".   The large extension of the CMEs with  spherical extended shock front  could explain
the long delay of  accelerated energetic particles particularly  when the source is close to the limb \citep{Kwon2018}.
The two short delay cases (\#1, \#10) (less than one hour) could be interpreted as due to magnetic reconnection in the flare sites. \citet{Kim2015} confirmed that the largest  SPE of 2002 (case \#1)  is related to proton accelerated at the flare site. Case \#10 has very close   onset times  for flare and CME. Therefore, it  is difficult to distinguish the SPEs coming from flare reconnection  from those due to  CME shocks.

However, the surprising cases  are  the two X-flare-halo CMEs related to flares   \#4 and \#5 with their sources near the disk center: they are related to multiple interacting CMEs with one halo CME which is a favorable case of geo-effectivity \citep{Lugaz2017}.
 But the ICMEs associated with  these flare-CME pairs have  also a weak southward orientation of B$_z^*$. {More generally,  low magnetic field strength enhancement indicates that} the ICME should have been deflected in the interplanetary medium. This is consistent with the recent review of \citet{Manchester2017} who wrote about   "all the different ways CMEs and ICMEs are rotated, reconfigured, deformed, deflected during their journey through the solar wind." 
 The  ICME-Earth geometry is  an important parameter   as demonstrated by \citet{Cho2017} after considering a single C-class flare leading to a large disturbance and a single  X-class flare leading to  weak geomagnetic effects. 
{ We  conclude that for the two X-flare-halo CMEs (\#4, \#5), satisfying all the favorable proxies  (source location, long flare duration, critical mass,  large kinetic energy) to forecast an intense geo-effective storm is not a sufficient condition. 
This demonstrates also that energetic flares with fast front side halo CMEs never lead to strong geomagnetic disturbances in 2002. Other parameters have to be taken into account (e.g., the direction of {B}, the B$_z^*$ value).}

 \subsection{Forecast of SEP - Flares}
We showed that the 12  X-class flares were mainly related to three large complex active region areas which produced recurrent  and long duration flares  in July and August, a  similar  scenario  that leads commonly to  extreme Space Weather events  (e.g.\ the Halloween events in 2003).
Moreover, we need to understand the magnetic configuration of the solar sources and their stability for forecasting a geo-effective event before observing the CME \citep{Torok2005,Ishiguro2017}. The magnetic topology of the active regions is a key parameter to understand if the flare will be followed by a halo CME which could be geo-effective \citep{Schrijver2007,Wang2007,Falconer2011,Zuccarello2017,Kontogiannis2019}. It is crucial to determine if the flare will be eruptive and if particles can escape as demonstrated by  \citet{Amari2018}. Large sunspot groups with a complex magnetic field such as in 2002 are not sufficient to predict geo-effective events (see Figure~\ref{source}). For example if the magnetic field above the active region is too strong  (e.g.\  a strong magnetic cage), the magnetic energy can only be released as thermal energy producing X-class flares without CME and without related geo-effective events (SEP or Dst) \citep{Mays2015,Thalman2015}.\\
Forecasts of energetic flares depend strongly on the amount of free (non-potential) magnetic field energy stored in an active region \citep{Kim2017,Guennou2017}. The free magnetic field energy depends on the twist and the entanglement of the magnetic field in an active region. The twist of the magnetic field lines corresponds to the magnetic helicity of an active region  \citep{Pariat2005,Demoulin2006,Dalmasse2015,Linan2018}.  In order to fully understand the chain from the Sun to the Earth, the partition of energy in the ARs should be evaluated. However it is difficult to forecast such a partition between kinetic energy (CME), thermal energy (X-ray flare) and  energetic particles (SPE). Computing the variation of the magnetic helicity automatically for each AR would give an indication of the onset time of CMEs   \citep{Nieves2019}. We need to better understand all the fundamental solar processes for making progress in deeper understanding the chain of the events. 
 
 \subsection{Forecast  of ICMEs}
The fast halo CMEs associated with our five X-class flares interact with a previous CME or  a streamer that can be identified by multiple shocks and sheaths in each case.
All these interactions should lead to geo-effective events  \citep{Lugaz2017}.  However, in 2002 most of the sources were close to the limb and we measured only the consequences of projection effects. 
The anomaly 
of flares occurring mostly at the limb in 2002 leads to the low dip of the geo-effectiveness that year. It is also true for the SPE events which could have been much more energetic and numerous if the flares had occurred from the disk central western part. 

The comparison between the existing catalogs revealed that \citet{Jian2006a} already identified the five ICMEs in our study. \citet{Richardson2010} identified the 2 cases \#4 and \#5.  \citet{Nieves2019} identified all the cases except case \#7.
In our study, we have identified  case \#6 which is not listed  in the any other catalogs. Case \#6 has  a backside  source (E95)  and is not  associated with an ICME  at L1 because of the travel  direction of the ejecta.
Still  the  front shock is clearly identified in the interplanetary medium and is associated with a minimum value of the Dst index.

For  cases 4 and 5  their two  fast halo CMEs (associated with
X-flares)  coming from the West part of the disk central region are associated
with ICMEs at L1 and the geoeffectivity of these events is weak. It
demonstrates that forecasting requires  either L1 measurements (but with
less than 1 hour time delay) {or ICME}  propagation models including
magnetic field.

{ However, the main unknown is the  interplanetary magnetic field  configuration during the travel of the ICME and how the ICME shape and direction is altered.  We have actually only in situ measurements at L1 given by Wind and ACE spacecraft.  Interaction between ICMEs and the heliospheric plasma sheet and current sheet during their propagation can significantly change the ICME/flux rope orientation at different  heliocentric distances \citep{Rodriguez2008,Winslow2016}.  Using Messenger, Venus express and ACE spacecraft it has been shown that the solar wind conditions do not recover after the passage of an ICME \citep{Janvier2019}.
The knowledge of the interplanetary medium will benefit from the next missions, namely the Parker Solar Probe and Solar Orbiter: we expect to learn more about the shapes of the flux-ropes  (magnetic cloud) in the heliosphere, their deformations and erosions with encounters in the solar wind, and their deflections.
An armada of small satellites turning around Mercure for example  could also provide very suitable in situ measurements.
For the forecasts, the numerical simulation models (like ENLIL - \citet{Odstrcil1999}) and EUHFORIA - \citet{Pomoell2018}) are based on measurements as well and are promising: the simulations compute  the interactions involving magnetic reconnection with co-rotating region structures in the solar wind which can alter ICMEs \citep{Winslow2016}. Recent simulations with EUHFORIA can even predict the arrival time of SPEs and  ICMEs  \citep{Scolini2018,Wijsen2019}.}

\section{Conclusion}
The present paper focus on the 12 X-class flares observed in 2002 and observations directly or indirectly related to them (CME, ICME, SPE, indices). With such a small data set, we cannot make proper statistics nor predictions. The novelty of our work is { in the study of the  details   of } the Sun-Earth chain of events related to all X-class flares over one year without pre-selection. The main results coming out of this study are that:

\begin{itemize}
    \item None of  the 12 events, even the ones associated with fast halo CMEs from the central disk region, and even with ICMEs and SPE, are geo-effective. It demonstrates that these fast halo CMEs, X-class flares, central disk region, ICMEs, SPEs are not good proxies for geo-effectiveness in 2002. 
    The case studies for the six cases with an associated  ICME at L1, coming from different solar longitudes, show for all of them  a weak integrated normalized $z-$component  of the interplanetary magnetic field, B$^*_z$. {bf $B_{z}^*$ is computed by the integration over time of the IMF $B_z$ component, divided by the total interval duration.} The weakness of  the IMF B$_z^*$ explains the very weak geo-effectiveness of these six cases. By the way, as shown \citet{Bocchialini2018}, B$_z^*$  is  more reliable than the commonly used B$_z$ peak value. The peak  and sign values are an instant picture of the IMF while B$_z^*$  is a more global parameter.
    A large  B$_z^*$ implies a long and intense reconnection process at the magnetopause, and increases the probability of having a strong magnetic storm, and a deep Dst decrease.
    \item For the six events  related to an ICME at L1, flare characteristics e.g.  longitude   and min(Dst) show a correlation  trend. Such a relationship between
    flares and min(Dst) would lead to long term prediction (1-3 days).
    However, due to our small  data set, we have to reproduce this study on a larger sample to have a robust conclusion.
    \item Finally, we offer a complete survey of 12 X-class flares related events in 2002. The only prediction we can make and explain is based on B$_z^*$  but L1 observations offer a very short window to raise alerts (less than 1h). The flare duration may give a larger window, but we would need to redo this study on a larger data set to check its robustness.
\end{itemize}
\bigskip
\bigskip
\begin{center}
{ANNEX}
\end{center}
{A. Halo CME sources in the West side (\#1 and \#10)}

Flare-CME pairs  \#1 and \#10 have a source  near the limb on the far  West side  which  could explain the weak effect on the Earth environment.  Each of them is related to an halo CME 
 followed by ICME and  SPE.
 These are the two most geo-effective events 
based on Dst and AE values (Table \ref{all_events}). It is worth noting that it corresponds to the two lowest B$_z^*$ values as well. For both events, the weak southward orientation of the IMF is the main explanation, as one can expect B$_z^*$ values lower than -5 nT to observe a stronger Dst depression \citep{Bocchialini2018}.
 Although flare \#1  has  a limb  source, the events associated with  this flare are  the largest ones  associated with an  X class flare-halo CME in  2002, 
{\it e.g.}, the largest CME speed, the largest SPE. However this flare-CME pair \#1 corresponds only to  a moderate magnetic storm.  For flare-CME pair  \#1 the particles are particularly energetic (2520 pfu) in the classification of SPEs.
Indices, SW, and IMF observations related to flare/CME \#10  have already been presented (see methodology, section \ref{methodo}).
The West side was favorable for having a relatively strong acceleration of particles in all the frequency ranges. However the SPE flux  for case \#10 is rather low.

\noindent  {B. Halo CME sources in the East side (\#6 and \#7)}
 
 The two  flare-CME pairs (\#6, \#7) in this group  associated with a halo CME  have an eastern source.
For flare/CME  \#6    (X3.3) on July 20, the source was  even backside  (S13, E95) but the SPE event  could be visible two days later  on July 21 at 07:00 UT (Figure \ref{spe_jul}). The SPE  event  lasts for 30 hours until July 22 12:00 UT. It is consistent with the results in  \citet{Cane1988} and \citet{Reames1999}. This gradual event with long duration from far East can be accounted for by IP shock acceleration  \citep{Cane1988}. 
\citet{Podgorny2018} explained that proton events coming from a source on the East side  was due to a transfer across the magnetic field lines with  the solar wind. 
Protons from West side are collisionless particles along the field lines and  a  delay of several hours  is expected.
The backside location (behind the eastern limb) of the solar source is also the main explanation for the lack of ICME observations at L1.

Flare-CME  pair \#7 is concomitant to a halo CME: with a  very well defined    shock front  (Figure \ref{case7}, see the arrows). 
Figure \ref{case7_L1} presents the in-situ measurements at L1 related to flare-CME   \#7.
Two discontinuities are observed in the magnitude of the IMF (top panel of Figure \ref{case7_L1}): 24/07 at 07:30 UT and 25/07 at 13:00 UT.
 \citet{Jian2006a} identified an ICME in between the two dashed lines (Figure \ref{case7_L1} top panel).
We now test the association between the discontinuities and the halo CMEs based on velocity comparisons.
As fast ICME shocks are progressively slowing down in the interplanetary medium at least up to 1 AU \citep{Gopalswamy2000,grison18}, we expect $v_{bal}$ to be lower than the CME velocity $v_{\odot{}}$ estimated from coronagraph observations and to be larger than the velocity observed at L1 ($v_{L1}$). This is the case for the discontinuity observed the 25 July 13:00 UT (flare-related CME \#6) ($v_{L1}= 520$;  $v_{bal}=625$; $v_{\odot{}}=2285$ km$\cdot$s$^{-1}$). The first discontinuity could be connected with the flare-CME  \#6.
The flare source was indicated as a backside source (S13, E90) but it could be just at the limb. The fast halo CME due to its spherical shape could reach L1 with an unclear ICME signature (indicated by "IP''  in Table \ref{all_events}). This velocity could reflect the projected velocity of the expansion of the halo CME or its spherical front  \cite{Kwon2018}. 

In addition to the B$_z$ IMF component -red curve in the second panel of Figure \ref{case7_L1}- the intensity of southward B$_z$ ($<0$) is displayed by a color scale in the top panel. Northward intervals are drawn in black and the strongest southward intervals are drawn in blue (B$_z < -5$ nT, see color scale). 
B$_z$ is the most southward  after 26/07 12:00 UT. This can explain why the minimum values of Dst is reached after that time (27/07 03:00 UT, cf. blue dot line in the bottom panel) and the weak response of the indices for that event.
For flare  \#7 no coordinated bump in the SPE records  around the time of the X4.8 class on July 23 at 00:42 UT was detected (Figure \ref{spe_jul}). Either the background was already too strong or the flare \#7 was not accompanied by any ejection of protons.  
 Only one - flare-CME pair  \#6  among the five  flare-CME pairs related to SPE events   is  on the East side of the Sun.   The  delay  between the flares and the particles is long, around 30  hours.  {This can be explained by 
 the East side location of the source favoring  not  a good magnetic connection.}
 
\noindent {C. Halo CME sources near the central meridian (\#4 and \#5)}

 The two flares \#4, \#5 and  {their associated  halo CMEs} have their sources  in the same  complex active region AR 10030 near the central meridian. The AR is    composed of two groups of sunspots (positive leading polarity, negative following polarity for each group) with a  strong interaction between them (Figure \ref{source}).
In fact flare-CME pairs   \#4 and \#5 are associated  with multi interactive CMEs \citep{Bocchialini2018}.  Associated with flare \#4 , {two CMEs  } have been identified in the LASCO field of view at 20:30 UT and at 21:30 UT.  Their kinetic energy has been estimated  to 1.6 $\times 10^{32}$ ergs (Table 2). Flare-CME pair \#5 is also  associated with two CMEs, a  partial  CME at 07:31 UT and the halo CME at 08:06 UT.
 
A convincing ICME signature is observed at L1 for these two events already analyzed by \citet{Richardson2010} and \citet{Bocchialini2018}.  
The interaction of CMEs leads commonly  to high SPE flux \citep{Lugaz2017}.
However the SPE flux is low and the value of the min(Dst) for each event indicates that the geomagnetic storm is also weak. The Dst is equal to  -17 nT  and -36 nT respectively (Table \ref{all_events}).
Conditions associated with flare-CME pairs  \#4 and \#5 (halo CME, fast CME speed,  multi CME interaction, ICME at L1, SPE)  are usually requested to lead to geo-effective phenomena.
We explain the low geo-effectiveness of these events (see AE and Dst columns in Table \ref{all_events}), once again with the weak southward IMF orientation (see B$_z^*$ value). Moreover IMF B$_z$ does not display a stable orientation in the ICME sheath following the shock.
The particles accelerated {  (event \#5) between  July 19  at 12:00 UT} and July 20 at 00 UT are coming from two  anti parallel directions   which could correspond to the two anchorage feet  of the ICME on the Sun. After the 20 July  00 UT, we note only particles at 0 degree  at L1 indicating that  only a part of the extreme flank of the ICME is crossed. This explains also the long sheath with  perturbed  B$_z$ (not shown).
The  delay  between the flares and the SPE particles for the two cases (flares \#4, \#5, \#6) are more than 20 hours. For these events the SPEs {are suspected } to be  released in ICME shocks.

\acknowledgments
B. Schmieder thanks Y.D.\ Park for his invitation of one month in KASI. We thank the two anonymous referees who help us to focus this study and arrive to a conclusion.
The authors  thank L.\ Klein for fruitful discussions on SPEs. This work was supported by the Programme National Soleil-Terre (PNST) of CNRS/INSU co-funded by CNES and CEA. B.\ Grison acknowledges support of the Czech Science Foundation grant 18-05285S and of the Praemium Academiae Award from the Czech Academy of Sciences.
Data analysis was performed with the AMDA science analysis system provided by the Centre de Donn\'{e}es de la Physique des Plasmas (CDPP) supported by CNRS, CNES, Observatoire de Paris and Universit\'{e} Paul Sabatier, Toulouse. The data used in this study were obtained from: \url{https://umbra.nascom.nasa.gov/SPE/} for the SPE list (NOAA/Space Weather Prediction Center), \url{http://omniweb.gsfc.nasa.gov/} for the events at L1 (OMNIWeb service), and the CDPP for the events at L1 and the geomagnetic indice \url{http://amda.cdpp.eu} (AMDA service). The list of flares and CMEs are available at \url{http://cdaw.gsfc.nasa.gov/CME\_list/}.

%
%
%

\end{document}